\documentclass[12pt]{article}
\usepackage{amsmath,amssymb,exscale,multirow,axodraw,color,epsfig,psfrag}
\input epsf
\newcommand{\m}[1]{\marginpar{{\tiny *}} }
\newcommand{\Dslash}{{\not \!\!D}}
\newcommand{\pslash}{{\not \!p}}
\newcommand{\tr}{{\rm tr}}

\DeclareMathOperator{\Diag}{Diag}

\begin{document}
\topmargin -1.0cm
\oddsidemargin -0.8cm
\evensidemargin -0.8cm

\thispagestyle{empty}

\vspace{40pt}

\begin{center}
\vspace{40pt}

\Large \textbf{Beautiful mirrors for a pNGB Higgs}

\end{center}

\vspace{15pt}
\begin{center}
{\large Eduardo C. Andr\'es, Leandro Da Rold, Iv\'an A. Davidovich} 

\vspace{20pt}

\textit{Centro At\'omico Bariloche, Instituto Balseiro and CONICET}
\\[0.2cm]
\textit{Av.\ Bustillo 9500, 8400, S.\ C.\ de Bariloche, Argentina}

\end{center}

\vspace{20pt}
\begin{center}
\textbf{Abstract}
\end{center}
\vspace{5pt} {\small \noindent
We consider one of the most significant deviations from the Standard Model: the forward-backward asymmetry of the $b$-quark measured at leptonic colliders. We investigate the possibility to solve this discrepancy by introducing new physics at the TeV scale. We focus on models where the Higgs is a pseudo Nambu-Goldstone boson of a new strongly coupled sector with a global SO(5) symmetry broken spontaneously to SO(4). Besides the usual top partners, we introduce bottom partners in the representations ${\bf 16}$ and ${\bf 4}$ of SO(5) and show that they can improve significantly the fit by correcting the $Zb\bar b$ couplings. We also estimate the corrections to the couplings at one-loop and obtain that the tree-level ones dominate and can give a reliable estimation. We find that the large shift required for $Zb_R\bar b_R$ leads to light custodians associated to the $b$-quark, similar to the top partners, as well as a rich phenomenology involving neutral interactions in the bottom-sector. 
}

\vfill\eject
\noindent

\section{Introduction}
The recent discovery of a light scalar field, with a mass around 126 GeV and similar properties to the Standard Model (SM) Higgs boson, has deep implications in our understanding of electroweak symmetry breaking (EWSB)~\cite{Aad:2012tfa,Chatrchyan:2012ufa}. The negative results in the search of new particles with masses $M\lesssim$ TeV at the first LHC-run introduces some tension for theories Beyond the Standard Model (BSM) aiming to solve the little hierarchy problem. One of the most interesting possibilities to alleviate this problem is a Higgs arising as a pseudo-Nambu Goldstone boson (pNGB) of a new strongly interacting sector at a few TeV scale~\cite{GK}. In this scenario the Higgs potential is generated radiatively through the interactions with the SM fields, that explicitly break the non-linearly realized global symmetry protecting the Higgs potential, leading to a separation between the scale of the resonances of the strongly coupled field theory (SCFT) and the electroweak (EW) scale. Besides, the contributions of the fermions to the potential, particularly those associated to the top, are misaligned with the original vacuum and can trigger EWSB. Although it has been shown that these theories still require some amount of tuning~\cite{Panico:2012uw}, they remain as one of the most promising BSM avenues.

There are different patterns of symmetry breaking that one can consider to obtain a Higgs as a pNGB. A very interesting approach is the minimal composite Higgs model (MCHM) based on SO(5)-symmetry. As first shown in Ref.~\cite{Agashe:2004rs}, SO(5)$\times$U(1)$_X$ is the minimal symmetry group that contains the EW gauge symmetry of the SM as well as a custodial symmetry and can deliver a Higgs as pNGB.~\footnote{The U(1)$_X$ factor is required to account for the proper normalization of hypercharge.} Assuming that the strong interactions of the SCFT spontaneously break SO(5) to SO(4), a Nambu Golstone boson (NGB) emerges with the proper quantum numbers to be associated with the SM Higgs. Several incarnations of the SO(5)/SO(4) symmetry breaking pattern have been considered in the literature. As examples, there are realizations in warped~\cite{Agashe:2004rs,Contino:2006qr} and flat extra-dimensions~\cite{Serone:2009kf,Pappadopulo:2013vca}, as well as theories with collective breaking or deconstruction, theories with two~\cite{DeCurtis:2011yx,Carena:2014ria} and three sites~\cite{Panico:2011pw}, that can be thought of as discretized descriptions of extra dimensional models. 

One of the main difficulties for composite Higgs models is to pass electroweak precision tests (EWPT). Generically these theories induce corrections to the SM interactions at tree level that can bee too large to pass these stringent tests. Besides the well known oblique corrections, one important problem is the correction to the $Zb_L\bar b_L$ coupling, that has been measured very precisely, such that the corrections can not be larger than the per mil level compared with the SM. In models where the BSM sector has a global O(4) symmetry spontaneously broken to the custodial O(3) after EWSB, it has been shown that the $Zb_L\bar b_L$ coupling can be protected by the presence of a subgroup of O(3). Assuming that the interactions between the SM and the BSM are linear in the SM fields, also known as partial compositeness:
\begin{equation}\label{pc}
{\cal L}_{\rm int}=\psi_{SM}{\cal O}^\psi_{\rm SCFT} \ , 
\end{equation}
this symmetry is realized if the SCFT operators coupled to the SM doublet $q_L$ transform as $({\bf 2},{\bf 2})_{\bf2/3}$ under SU(2)$_L\times$SU(2)$_R\times$U(1)$_X$~\cite{Agashe:2006at}. This way, despite the large mixing between the resonances of the SCFT and $q_L$, required to obtain the top mass, to leading order there are no corrections to $Zb_L\bar b_L$. Moreover, when considering a pNGB Higgs arising from SO(5), the previous assignment of quantum numbers restricts the SO(5) representations of the operator ${\cal O}^{q}_{\rm SCFT}$ to include a $({\bf 2},{\bf 2})$, the smallest representations satisfying this condition are ${\bf 5}$, ${\bf 10}$ and ${\bf 14}$~\cite{Agashe:2006at,Contino:2006qr,Panico:2012uw}.

On the other hand, LEP and SLD measurements of the forward-backward asymmetry in the production of $b\bar b$: $A^b_{FB}$, suggest deviations of the coupling $Zb_R\bar b_R$ compared with the SM. A solution to this anomaly can be achieved by the introduction of new vector-like fermions with the proper quantum numbers, often called {\it beautiful mirrors}~\cite{Choudhury:2001hs}. In the framework of composite Higgs models, these fermions can be associated to excitations created by the operator ${\cal O}^{b}_{\rm SCFT}$. Ref.~\cite{Agashe:2006at} showed the expected sign in the shift of the $Zb_R\bar b_R$ coupling for some representations of ${\cal O}^{b}_{\rm SCFT}$ under SU(2)$_L\times$SU(2)$_R\times$U(1)$_X$. Refs.~\cite{DaRold:2010as,Alvarez:2010js} proposed an effective two-site model with two resonances mixing with the bottom sector, one transforming as $({\bf 1},{\bf 2})_{-5/6}$ that induces the proper correction to $Zb_R\bar b_R$, and another one transforming as $({\bf 2},{\bf 3})_{-5/6}$ that after small mixing with $b_L$ gives a small positive correction to $Zb_L\bar b_L$, preferred by $A^b_{FB}$. The presence of both multiplets also allows to write a proto-Yukawa term in the sector of resonances, that can lead to the mass of the bottom quark through partial compositeness. Refs.~\cite{Djouadi:2006rk,Bouchart:2008vp} have proposed similar solutions.

Recently Ref.~\cite{Ciuchini:2014dea} performed a fit of the SM taking into account the latest theoretical and experimental results. It shows significant deviations from the SM in $Zb\bar b$ couplings that are attributed to deviations in $A^b_{FB}$ at $2.8\sigma$. The fit  points to $\delta g/g$ corrections of ${\cal O}(20\%)$ for the Right-handed coupling correlated with a $\sim1\%$ correction of the Left-handed one.

Motivated by the discovery of a light Higgs-like particle and the deviations in the $Zb\bar b$ couplings, we want to consider an extension of the model of Ref.~\cite{DaRold:2010as} where the Higgs could be realized as a pNGB. We will consider an effective two site model with the Higgs arising from the well known SO(5)/SO(4) pattern of symmetry breaking. To obtain a finite one-loop potential we will embed all the fermion and vector resonances into full SO(5) multiplets~\cite{DeCurtis:2011yx}. To generate the proper corrections of the $Zb\bar b$ couplings we will consider bottom partners embedded in SO(5) representations containing a $({\bf 1},{\bf 2})$ and a $({\bf 2},{\bf 3})$ multiplet of SU(2)$_L\times$SU(2)$_R$. The smallest representations of SO(5) with these properties are the {\bf 4} and {\bf 16}; these will be the uplifted beautiful mirrors. To be able to generate the top mass and trigger EWSB we will also introduce top partners. For simplicity we will consider them transforming with the representation {\bf 5} of SO(5).

In sec.~\ref{model} we describe our model and compute the $Zb\bar b$ couplings at tree level in the mass basis. 
In sec.~\ref{leff} we show the effective theory at energies much lower than the scale of the {\it composite}-resonances and obtain the couplings at zero momentum. 
In sec.~\ref{higgspotential} we compute the Higgs potential arising from the specified set of representations. 
In sec.~\ref{Zcouplingsnumeric} we give our numerical results after demanding dynamical EWSB as well as the proper spectrum for the light states corresponding to the SM degrees of freedom. 
In sec.~\ref{Zcouplingsloop} we discuss the one-loop corrections to the $Zb\bar b$ couplings.
In sec.~\ref{pheno} we give a brief description of some interesting properties of the model for the phenomenology at accelerators and we conclude in sec.~\ref{conclusions}.

\section{A model to solve the deviation in $A_{FB}^b$ and a light Higgs}\label{model}
Ref.~\cite{Agashe:2006at} considered theories with a new SCFT with global symmetry SU(2)$_L\times$SU(2)$_R\times$U(1)$_X$. It showed that if the SM fields interact linearly with the SCFT, Eq.~(\ref{pc}), in order to protect the $Zb_L\bar b_L$ coupling from large corrections in the presence of a large mixing between $q_L$ and an operator ${\cal O}_{SCFT}^{q}$, the new sector must have a $P_{LR}$ symmetry exchanging SU(2)$_L$ and SU(2)$_R$, and ${\cal O}_{SCFT}^{q}$ must be embedded in the $({\bf 2},{\bf 2})_{\bf 2/3}$ representation of the extended symmetry. In this case the interaction between $b_L$ and the corresponding component of ${\cal O}_{SCFT}^{q}$ preserves $P_{LR}$ and the coupling $Zb_L\bar b_L$ is protected. On the other hand Refs.~\cite{Agashe:2006at} and~\cite{DaRold:2010as} showed that $\delta g_{b_R}$ can be positive if $b_R$ mixes with an ${\cal O}_{SCFT}^{b}$ in a $({\bf 1},{\bf 2})_{\bf -5/6}$ and~\cite{DaRold:2010as} also showed that $\delta g_{b_L}$ can be positive if $q_L$ mixes with a second operator ${\cal O}_{SCFT}^{q^b}$ in a $({\bf 2},{\bf 3})_{\bf -5/6}$. 

By extending the symmetry of the new sector to SO(5)$\times$U(1)$_X$, the Higgs can be a pNGB arising from the spontaneous breaking of SO(5) to SO(4). The interactions between the SCFT and the SM fields explicitly break the global symmetry and generate a potential for the Higgs at loop level. This potential can induce a misalignment of the vacuum and trigger EWSB dynamically. Explicit realizations of this pattern of symmetry breaking \cite{Agashe:2004rs,Carena:2006bn,Contino:2006qr,Carena:2007ua,DeCurtis:2011yx,Panico:2011pw,Carena:2014ria} have shown that to obtain a finite Higgs potential at one loop the ${\cal O}_{SCFT}$ must be embedded in full representations of the global symmetry. In the present work we are interested in the representations {\bf 16}  and {\bf 4} of SO(5) for the fermionic operators ${\cal O}_{SCFT}^{q^b}$ and ${\cal O}_{SCFT}^{b}$, respectively, since they contain a $({\bf 2},{\bf 3})$ and $({\bf 1},{\bf 2})$ multiplets of SU(2)$_L\times$SU(2)$_R$, and therefore the interactions of this operators with the bottom quark can induce the proper shifts of the $Zb\bar b$ couplings to solve the deviations pointed to in Ref.~\cite{Ciuchini:2014dea}. These operators also have the proper quantum numbers to generate a bottom Yukawa interaction. 

There are larger SO(5)-representations containing the multiplets of SU(2)$_L\times$SU(2)$_R$ specified previously and there are also larger SU(2)$_L\times$SU(2)$_R$-representations that can achieve shifts in the $Zb\bar b$ couplings with the correct sign~\cite{DaRold:2010as}, however the ${\bf 4}$ and ${\bf 16}$ are the smallest representations once SO(5) is chosen; in this sense we will work in a minimal model. These operators can create fermionic resonances, the lowest lying levels of those with masses of order TeV, that we will call beautiful mirrors.

To be able to generate the top mass by partial compositeness and trigger EWSB we also introduce an operator ${\cal O}_{SCFT}^{q^t}$ in a representation of SO(5), we use the index $q^t$ to distinguish it from the one involved in the generation of the bottom mass that has index $q^b$. We will add an operator ${\cal O}_{SCFT}^{t}$ interacting with $t_R$, and for simplicity we will assume that ${\cal O}_{SCFT}^{q^t}$ and ${\cal O}_{SCFT}^{t}$ transform with the representation ${\bf 5}$. We could choose other representations for these operators, but we do not expect that the physics we want to study could have a relevant dependence on this choice.

\subsection{2-site description}
A 4D effective description of an SCFT leading to resonances, similar to QCD but with a scale of order TeV, can be realized in a theory with two sites~\cite{Contino:2006nn}: one site called site-0 with {\it elementary} fields and another one called site-1 with fields that describe, at an effective level, the first layer of {\it composite} resonances of the SCFT. In this work we will consider a two site theory very similar to the models of Refs.~\cite{DeCurtis:2011yx} and~\cite{Carena:2014ria}, with the difference that we will introduce representations for the fermions not considered in those articles. We will closely follow the notation of Ref.~\cite{Carena:2014ria}.

We will use lower case letters for the fields of site-0 and capital letters for the fields of site-1.
The {\it elementary} sector at site-0 contains the same fermionic and gauge degrees of freedom as the SM but no elementary scalar:
\begin{equation}
{\cal L}_0=-\frac{1}{4g_0^2}\tilde w^j_{L\mu\nu}\tilde w^{j\ \mu\nu}_L
-\frac{1}{4g_0^{'2}}\tilde b_{\mu\nu}\tilde b^{\mu\nu} + i\bar\psi\Dslash_0\psi \ ,
\end{equation}
where a sum over the SM fermions is understood, $j=1,2,3$ and $\tilde w^j_L$ and $\tilde b$ are the SU(2)$_L$ and U(1)$_Y$ gauge fields, respectively. $D^\mu_0$ is the covariant derivative containing the fields of the {\it elementary} sector. The tilde over the gauge fields denotes the non-canonical normalization of their kinetic terms. There is also an SU(3)$_c$ gauge symmetry that we have not written because it does not play any role in our analysis.

It is very convenient to introduce new spurious degrees of freedom in the {\it elementary} sector~\cite{Agashe:2004rs}, such that the SM gauge symmetry is extended to SO(5)$_0\times$U(1)$_x$, the same group as in the {\it composite} sector, with hypercharge realized as $Y=T^{3R}+X$.~\footnote{For more information on the representations of SO(5) used in this paper see Ap.~\ref{app-reps}.} The spurious fields are not dynamical and they do not play any physical role. The fermions can also be extended to fill complete representations of this group. 
We will introduce two chiral {\it elementary} fermions called $q^t$ and $q^b$, respectively transforming as ${\bf 5_{2/3}}$ and ${\bf 16_{-5/6}}$. Under SU(2)$_L\times$SU(2)$_R$ these multiplets decompose as: ${\bf 5}\sim ({\bf 2},{\bf 2})\oplus({\bf 1},{\bf 1})$ and ${\bf 16}\sim ({\bf 3},{\bf 2})\oplus({\bf 2},{\bf 3})\oplus({\bf 2},{\bf 1})\oplus({\bf 1},{\bf 2})$. The $({\bf 2},{\bf 2})_{\bf 2/3}$ and the $({\bf 2},{\bf 3})_{\bf -5/6}$ contain each just one ${\bf 2_{1/6}}$ of SU(2)$_L\times$U(1)$_Y$, whereas the other SO(4) multiplets do not contain a ${\bf 2_{1/6}}$. Out of these two doublets only one linear combination will be dynamical. By defining $P_{\bf 2_{1/6}}$ as the projector that acting on the space of the fields embedded in full SO(5) multiplets project them onto the subspace containing a ${\bf 2_{1/6}}$, the dynamical doublet can be written as $q_L\equiv P_{\bf 2_{1/6}}(q^t+q^b)$, all the other components will be just spurious fields. We will also promote $t_R$ to a ${\bf 5_{2/3}}$, with the $({\bf 1},{\bf 1})_{\bf 2/3}$ component being the only dynamical field. We will embed $b_R$ in a ${\bf 4_{-5/6}}$, using that ${\bf 4}\sim ({\bf 2},{\bf 1})\oplus({\bf 1},{\bf 2})$, the only dynamical field will be the up component of the $({\bf 1},{\bf 2})_{\bf -5/6}$. In terms of these multiplets the {\it elementary} Lagrangian reads:
\begin{equation}
{\cal L}_0=-\frac{1}{4g_0^2}\tilde a^C_{\mu\nu}\tilde a^{C\mu\nu}
-\frac{1}{4g_{x}^{2}}\tilde x_{\mu\nu}\tilde x^{\mu\nu} + i\bar\psi\Dslash_0\psi \ ,
\end{equation}
where now the sum is over $\psi=q^t,q^b,t$ and $b$, $C$ is an index in the adjoint of SO(5): $C=1,\dots 10$. The {\it elementary} hypercharge coupling is $g_0^{\prime-2}=g_0^{-2}+g_x^{-2}$ and the dynamical field is obtained by setting $\tilde w^3_R=\tilde x=\tilde b$. 

The {\it composite} sector at site-1 contains an SO(5)$_1\times$U(1)$_X$ gauge symmetry, as well as several fermions charged under this symmetry.\footnote{One can consider also an SU(3) gauge symmetry in the {\it composite} sector to describe resonances of the gluons. For simplicity, and because we will be concerned with the EW sector only, we will not mention them anymore in this work.} We will assume that the SO(5)$_1$ symmetry is spontaneously broken to SO(4)$_1$ at a scale $f_1$ by the strong dynamics of site-1. We will parametrize this breaking by considering a non-linear description in terms of a unitary matrix $U_1$, containing the NGB fields arising from the spontaneous breaking: $U_1=e^{\sqrt{2}i\Pi_1/f_1}$, with $\Pi_1=\Pi_1^{\hat a}T^{\hat a}$ and $T^{\hat a}$ the broken generators of SO(5)$_1$/SO(4)$_1$. The SO(5)$_1$ symmetry is non-linearly realized, since under a transformation at site-1: ${\cal G}_1 \in{\rm SO(5)}_1$: $U_1 \to {\cal G}_1 \, U_1 \, {\cal H}_1({\cal G}_1;\Pi_1)^\dagger$, with ${\cal H}_1({\cal G}_1;\Pi_1) \in {\rm SO(4)}_1$ implicitly depending on ${\cal G}_1$ and $\Pi_1$, as usual in the NGB formalism~\cite{CCWZ}. The Lagrangian of the bosonic sector at site-1 is:
\begin{equation}
{\cal L}_1\supset-\frac{1}{4g_\rho^2}\tilde A^C_{\mu\nu}\tilde A^{C\mu\nu}
-\frac{1}{4g_X^{2}}\tilde X_{\mu\nu}\tilde X^{\mu\nu}+\frac{f_1^2}{2}{\cal D}_\mu^{\hat a}{\cal D}^{\mu\hat a} 
\end{equation}
with $C=1,\dots 10$ and $\tilde A$ and $\tilde X$ being the SO(5)$_1$ and U(1)$_X$ gauge fields, respectively. ${\cal D}^{\hat a}_\mu$ is implicitly defined by $U_1^\dagger D_{1\mu} U_1=i{\cal E}^{a}_\mu T^a+{\cal D}^{\hat a}_\mu T^{\hat a}$, with $D_{1\mu}$ the covariant derivative containing the fields of the {\it composite} sector. We take the couplings of site-1 to be larger than the SM couplings, but still in the perturbative regime: $g_{SM}\ll g_\rho\ll 4\pi$, where by $g_\rho$ we generically denote all the couplings of the {\it composite} sector.

For the fermions of the {\it composite} sector we will consider four vector-like multiplets: two associated to the top and transforming as ${\bf 5}_{2/3}$, called $Q^t$ and $T$, and two associated to the bottom, one called $Q^b$ transforming as ${\bf 16}_{-5/6}$ and another one called $B$ transforming as ${\bf 4}_{-5/6}$. Besides the usual kinetic and mass terms, there are also Yukawa interactions involving the NGB field $\Pi_1$. The Lagrangian of the fermions at site-1 is:
\begin{equation}\label{Lcpfermion}
{\cal L}_1\supset \bar\Psi(i\Dslash_1-m_\Psi)\Psi+\sum_ry_{tr}{P_r(\overline{U_1^\dagger Q^t_L})}P_r(U_1^\dagger T_R)+\sum_ry_{br}P_r(\overline{U_1^\dagger Q^b_L})P_r(U_1^\dagger B_R)+{\rm h.c.}
\end{equation}
where a sum over $\Psi=Q^t,Q^b,T$ and $B$ is understood. $r$ is an irreducible representation of SU(2)$_L\times$SU(2)$_R$, $P_r$ is a projector from the space of representations of SO(5) to the subspace of the $r$ representation of SU(2)$_L\times$SU(2)$_R$ and the product $P_r(\Phi')P_r(\Phi)$ corresponds to the usual operation leading to an SU(2)$_L\times$SU(2)$_R$ invariant.~\footnote{It is also usual to write the invariants by working with the NGB vector field $\Phi_1=U_1\Phi_0$, with $\Phi_0^t=(0,0,0,0,1)$ parametrizing the SO(5)/SO(4) vacuum before EWSB. In this case the invariants for the Yukawa interactions can be written formally as: $\bar \Psi_L\Phi^n\Psi'_R$, with $n=2$ for $\Psi,\Psi'\sim{\bf 5}$ and $n=1,2$ for $\Psi\sim{\bf16}$ and $\Psi'\sim{\bf4}$. Refs.~\cite{DeCurtis:2011yx,Carena:2014ria} used the later parametrization to write the Yukawa interactions.} We have considered only a partial set of chiral structures, not including for example terms of the form $P_r(\overline{U_1^\dagger \Psi_L})P_r(U_1^\dagger \Psi_R)$, neither of the form: $P_r(\overline{U_1^\dagger Q^t_R})P_r(U_1^\dagger T_L)$ or $P_r(\overline{U_1^\dagger Q^b_R})P_r(U_1^\dagger B_L)$. As argued in Ref.~\cite{Carena:2014ria}, those operators introduce divergences in the Higgs potential at one-loop, unless one goes to three or more sites~\cite{DeCurtis:2011yx,Panico:2011pw}. For the Yukawa interactions of the top sector there is a trivial singlet, independent of $\Pi_1$~\cite{Mrazek:2011iu}, that leads to a mass mixing term between $Q^t$ and $T$ and can be obtained by taking $y_{t({\bf 1},{\bf 1})}=y_{t({\bf 2},{\bf 2})}$. There is also a non-trivial invariant proportional to $y_{t({\bf 1},{\bf 1})}-y_{t({\bf 2},{\bf 2})}$. For the bottom sector there are two non-trivial independent invariants, with couplings $y_{b({\bf 1},{\bf 2})}$ and $y_{b({\bf 2},{\bf 1})}$. The presence of more than one Yukawa structure in the down sector in general leads to flavor violating processes mediated by Higgs exchange that, in anarchic models, are too large compared with the present bounds on flavor violating interactions~\cite{Agashe:2009di}. By imposing a $P_{LR}$ symmetry under the exchange of SU(2)$_L$ and SU(2)$_R$ one obtains: $y_{b({\bf 1},{\bf 2})}=y_{b({\bf 2},{\bf 1})}$. In a theory of flavor this symmetry aligns the bottom Yukawa structures in flavor space and relaxes the most stringent constraints arising from those processes~\cite{Agashe:2009di}. In the following we will assume that $y_{b({\bf 1},{\bf 2})}=y_{b({\bf 2},{\bf 1})}\equiv y_b$. Notice that the Yukawa couplings as defined in Eq.~(\ref{Lcpfermion}) are dimensionful. We expand the neutral Yukawa interactions in terms of the components within each multiplet in Ap.~\ref{App-yukawa}. 

The {\it elementary} and {\it composite} sectors described above are coupled by non-linear $\sigma$-model fields $\Omega$ and $\Omega_X$, realizing partial compositeness. The non-linear $\sigma$-model fields transform bilinearly under the {\it elementary} and {\it composite} symmetry groups: $\Omega\to g_0\Omega g_1^\dagger$ and similarly for $\Omega_X$. As a consequence, there is mixing between the {\it elementary} and {\it composite} fields, and the mass eigenstates are a superposition of both sectors. Each non-linear $\sigma$-model field is characterized by an energy scale, $f_\Omega$ and $f_{\Omega_X}$. The mixing Lagrangian is:
\begin{align}
{\mathcal L}_{\rm mix} = &\frac{f_\Omega^2}{4}\tr|D_\mu\Omega|^2 + \frac{f_{\Omega_X}^2}{4} |D_\mu\Omega_X|^2 
+\bar q_L\Omega \, (\Delta_{Q^t}\Omega_X^{2/3}Q^t_R+\Delta_{Q^b}\Omega_X^{-5/6}Q^b_R)
\nonumber \\
&+\Delta_T\bar t_R\Omega \Omega_X^{2/3}T_L+\Delta_B\bar b_R\Omega \Omega_X^{-5/6}B_L+{\rm h.c.}
\end{align}
The link field $\Omega$ parametrizes the coset ${\rm SO(5)}_0 \times {\rm SO(5)}_1 / {\rm SO(5)}_{0+1}$, with SO(5)$_{0+1}$ the diagonal subgroup of ${\rm SO(5)}_0 \times {\rm SO(5)}_1$, and similarly for $\Omega_X$
\begin{align}
\Omega = e^{\sqrt{2} \, i \Pi_{\Omega} / f_\Omega}~,
\qquad
\Omega_X = e^{\sqrt{2} \, i \Pi_{\Omega_X} / f_{\Omega_X}}~,
\end{align}
where $\Pi_{\Omega} = \Pi_{\Omega}^{C} \, T^C_{0-1}$ and $T^C_{0-1}$ are the generators of the coset. The covariant derivatives contain in this case {\it elementary} and {\it composite} gauge fields, as required by gauge invariance:  
\begin{equation}
D_\mu\Omega = \partial_\mu \Omega - i \tilde{a}_\mu\Omega + i\Omega \tilde{A}_\mu \ ,
\qquad
D_\mu\Omega_X = \partial_\mu \Omega_X - i \tilde{x}_\mu\Omega_X + i\Omega_X \tilde{X}_\mu \ .
\end{equation}

\subsection{Mass eigenstate basis}
To make contact with the physical content of the theory one can go to the unitary gauge by performing a gauge transformation ${\cal G}_1=\Omega$ and ${\cal G}_{1X}=\Omega_X$. After that the only physical scalars can be parametrized by
\begin{equation}
U=e^{\sqrt{2} \, i \Pi / f_h} \ ,
\qquad
\Pi=h^{\hat a}T^{\hat a} \ ,
\qquad
\frac{1}{f_h^2}=\frac{1}{f_\Omega^2}+\frac{1}{f_1^2} \ .
\end{equation}
After EWSB the vacuum is described by the parameter:
\begin{equation}
\epsilon=\sin\frac{v}{f_h} \ ,
\end{equation}
with $v=\langle h\rangle$ and $h^2=h^{\hat a}h^{\hat a}$.

To better understand the particle content and the different sources of symmetry breaking of the theory, we briefly discuss the spectrum  as well as its modifications when the different sources of mixing are taken into account. Freezing the {\it elementary} fields one can study the spectrum of the pure {\it composite} sector. It contains vectors in the SO(4)$_1$ subgroup with mass $m_\rho=g_\rho f_\Omega/\sqrt{2}$, vectors in the SO(5)$_1/$SO(4)$_1$ coset with mass $m_{\hat a}=g_\rho \sqrt{f_\Omega^2+f_1^2}/\sqrt{2}$ and a vector of U(1)$_X$ with mass $m_X=g_Xf_{\Omega_X}/\sqrt{2}$.

After the mixing between the two sites, there remains a gauge symmetry SU(2)$_{L,0+1}\times U(1)_{Y,0+1}$, corresponding to the diagonal subgroups. The corresponding massless gauge fields are linear combinations of {\it elementary} and {\it composite} gauge fields
\begin{equation}\label{massless-gauge}
W_L^i=c_\theta\ w_L^i+s_\theta A_L^i \ ,\quad
i=1,2,3 \ ;\qquad
B=\frac{b+t_{\theta'_\rho}A_R^3+t_{\theta'_X}X}{(1+t_{\theta'_\rho}^2+t_{\theta'_X}^2)^{1/2}} \ ,
\end{equation}
where we have rescaled the gauge fields as: $w_L^i=g_0\tilde w_L^i$, $b=g'_0\tilde b$, $A^C=g_\rho\tilde A^C$ and $X=g_X\tilde X$. We have also used shorthand for the trigonometric functions: $c_\alpha\equiv\cos\alpha$, $s_\alpha\equiv\sin\alpha$ and $t_\alpha\equiv\tan\alpha$, and the different mixing angles: $\theta, \theta'_\rho$ and $\theta'_X$ are respectively given by: 
\begin{equation}
t_\theta=g_0/g_\rho \ , \quad t_{\theta'_\rho}=g'_0/g_\rho \ , \quad t_{\theta'_X}=g'_0/g_X \ . 
\end{equation}
The couplings of the gauge fields defined in Eq.~(\ref{massless-gauge}) are given by:
\begin{align}\label{matching-massless}
\frac{1}{g^2}=\frac{1}{g_0^{2}}+\frac{1}{g_\rho^2}\ ,
\qquad
\frac{1}{{g'}^2}=\frac{1}{g_0^{'2}}+\frac{1}{g_\rho^2}+\frac{1}{g_X^2} \ .
\end{align}
The orthogonal combinations to Eq.~(\ref{massless-gauge}) correspond to massive fields. The spectrum of heavy vectors arising from this diagonalization is slightly modified with respect to the original mass, but the corrections are small as long as $t_\theta\equiv g_0/g_\rho\ll 1$, as was assumed to be the case. As an example, for $\tilde\rho_L^i=c_\theta A_L^i-s_\theta w_L^i$ the mass is $m_{\tilde\rho}=m_\rho\sqrt{1+t_\theta^2}$. The masses of the {\it composite} fields that do not mix with the {\it elementary} ones remain unchanged. For more details on the precise linear combinations and spectrum of the massive states we refer the reader to Ref.~\cite{Carena:2014ria}.

After EWSB the spectrum is further modified, the $W$ and $Z$ bosons obtain masses: $m_Z\simeq \sqrt{g^2+{g'}^2}\epsilon f_h/2$ and $m_W\simeq g\epsilon f_h/2$, where one can identify:
$v_{SM}=246\ {\rm GeV}\simeq\epsilon f_h \ .$
The masses of the heavy states are slightly modified after EWSB, which induces mixing between the heavy states as well as mixing with the light ones. As a summary, there are seven neutral states: one massless and another light one, corresponding to the photon and $Z$, and five heavy states; there are also four charged states: a light one corresponding to the $W$, and three heavy states. Since the mixing angles are small, the light states are mostly {\it elementary} and the heavy states are mostly {\it composite}.

The fermionic mixing can also be diagonalized by performing a rotation of the chiral components involved in ${\cal L}_{\rm mix}$. We define fermionic mixing $t_{\theta_\Psi}\equiv\tan\theta_\Psi=\Delta_\Psi/m_\Psi$, with $\Psi=Q^t,Q^b,T,B$. After the corresponding rotation there is a set of chiral fermions that remain massless, their degree of compositeness measured by $\theta_\Psi$. The masses of the fermions corresponding to the orthogonal combinations become $m_{\tilde \Psi}=m_\Psi\sqrt{1+t_{\theta_\Psi}^2}$.~\footnote{Since $q_L$ mixes simultaneously with one doublet contained in $Q^t$ and another doublet contained in $Q^b$, the diagonalization of this system is more involved, requiring the angles $\theta_{Q^t}$ and $\theta_{Q^b}$. The squared masses of the massive states emerging from the diagonalization, in the absence of Yukawa terms, can be written in the following way: $\frac{1}{2}(m_{\tilde{Q^t}}^2+m_{\tilde{Q^b}}^2)\pm\frac{1}{2}[(m_{\tilde{Q^t}}^2+m_{\tilde{Q^b}}^2)^2-4m_{\tilde{Q^t}}^2m_{\tilde{Q^b}}^2(1+t_{\theta_{Q^t}}^2+t_{\theta_{Q^b}}^2)(1+t_{\theta_{Q^t}}^2)^{-1}(1+t_{\theta_{Q^b}}^2)^{-1}]^{1/2}$.} 
The masses of the fermions that do not mix, often called custodians, remain being $m_\Psi$. For large mixing angles, as is the case for the mixing leading to the top mass and mildly for the mixing of $b_R$ leading to the shift in the coupling $Zb_R\bar b_R$, there can be a sizeable separation between the scales $m_{\tilde \Psi}$ and $m_\Psi$. By fixing the scale $m_{\tilde \Psi}\sim m_{\tilde\rho}\sim{\cal O}$(2-3)TeV, for mixing $s_{\theta_\Psi}\gtrsim 0.7$, one obtains light custodians with masses $\lesssim1$ TeV~\cite{Agashe:2004ci,Contino:2006qr}. After EWSB all of the fermions with the same electromagnetic charge are mixed. In the present model there are nine up-type fermions, eleven down-type fermions and some exotic fermions: two with $Q=+5/3$, eight with $Q=-4/3$ and two with $Q=-7/3$. One up-type fermion and one down-type fermion become massive only after EWSB, they are lighter than the other states and correspond to the top and bottom. In Ap.~\ref{App-yukawa} we show the mass matrices for these fermions.

Although the mass matrices can be diagonalized straightforward numerically, it is worth obtaining the analytic diagonalization expanding in some small parameter. By this procedure one can obtain analytic expressions for the couplings in the mass basis and understand the size and sign of the corrections as functions of the fundamental parameters of the theory. We have considered a perturbative expansion in powers of $\epsilon$, obtaining a full diagonalization of the fermionic and bosonic sectors to ${\cal O}(\epsilon^2)$. Since the mass matrices of the up- and down-type quarks are of dimension 9 and 11, and the mass matrix of the neutral vector bosons is of dimension 7 \cite{Carena:2014ria}, the explicit expressions for the eigenvalues and eigenvectors are too long to be written in the paper. However in the next subsection we will show our results for the $Zb\bar b$ interactions using this perturbative diagonalization.

\subsection{$Zb\bar b$ interactions in the mass eigenstate basis}\label{Zcouplingsmassbasis}
We compute the $Zb\bar b$ interactions in the basis of mass eigenstates. In order to do that we have diagonalized the bosonic and fermionic mass matrices perturbatively in $\epsilon$, the first non-trivial correction being of ${\cal O}(\epsilon^2)$.

Since the $Z$ is a mixing between the {\it elementary} and {\it composite} neutral vector fields, there is a universal correction to the $Z$ couplings. That correction remains for vanishing fermion mixing and therefore is the same for all fermions. We have subtracted that term in the results that we present in Eq.~(\ref{eqZcouplingsmass}), such that we only show the non-universal corrections. In order to simplify the equations, for the correction to $Zb_L\bar b_L$ we have set $\Delta_B=0$, whereas for $Zb_R\bar b_R$ we have set $\Delta_{Q^b}=0$, obtaining:
\begin{align}\label{eqZcouplingsmass}
\delta g^{\rm mass}_{b_L}=&\frac{g}{c_W}\ \frac{1}{2}\ \frac{\epsilon^2f_h^2}{m_{Q^t}^2m_T^2\Delta_{Q^b}^2+m_{Q^b}^2[m_{Q^t}^2m_T^2+\Delta_{Q^t}^2(m_T^2+y_{t({\bf2},{\bf2})}^2)]}\nonumber \\
&\left\{-\Delta_{Q^t}^2m_{Q^b}^2(m_T^2+y_{t({\bf2},{\bf2})}^2)\left[\frac{g^2-{g'}^2}{4m_\rho^2}+\frac{{g'}^2}{3m_X^2}\right]\right. \nonumber \\
&\ \left.+\Delta_{Q^b}^2 m_{Q^t}^2 m_T^2\left[\frac{g^2(2+3t_\theta^{-2})-2{g'}^2}{4m_\rho^2} + \frac{5{g'}^2}{12m_X^2}+\frac{5y_{b({\bf 1},{\bf 2})}^2}{8f_h^2m_B^2}\right]\right\}\ ,
\nonumber \\
\delta g^{\rm mass}_{b_R}=&\frac{g}{c_W}\ \frac{\epsilon^2f_h^2}{m_{Q^b}^2m_B^2+\Delta_B^2(m_{Q^b}^2+y_{b({\bf1},{\bf2})}^2)}\ \Delta_B^2\nonumber \\
&\left\{\frac{5{g'}^2}{24m_X^2}(m_{Q^b}^2+y_{b({\bf1},{\bf2})}^2)+\frac{y_{b({\bf 1},{\bf 2})}^2+y_{b({\bf 2},{\bf 1})}^2}{8f_h^2}\right. \nonumber \\
&\left. \ +\frac{1}{8m_\rho^2}\left[g^2(1+t_\theta^{-2})(m_{Q^b}^2+2y_{b({\bf1},{\bf2})}^2)-{g'}^2(m_{Q^b}^2+y_{b({\bf1},{\bf2})}^2)\right] \right\}\ .
\end{align}
The gauge couplings appearing in Eq.~(\ref{eqZcouplingsmass}) were defined in Eq.~(\ref{matching-massless}). We have checked that the relative difference between these approximations and the full numerical results are $\sim1\%$ for the points selected in our scan (see details in the next sections). We analyse first the Left-coupling. Notice that, when considering the interaction between mass-eigenstates, this coupling is modified by the mixing with the fermion $Q^t$ even for $\Delta_{Q^b}=0$, the correction being of order $\delta g/g\sim -g^2\epsilon^2\Delta_{Q^t}^2/4g_\rho^2m_{Q^t}^2$. Compared with the naive estimate in the absence of $P_{LR}$-symmetry~\cite{Giudice:2007fh}, this contribution is suppressed by a factor $g^2/g_\rho^2$. This term is present because the mass eigenstates are mixtures of {\it elementary} and {\it composite} fields, therefore they do not have well defined transformation properties under the full gauge symmetry group. The correction arising from the mixing with $Q^b$ can be split in two: one mediated by a heavy vector, of order $\delta g/g\sim \epsilon^2\Delta_{Q^b}^2/m_{Q^b}^2$, where we have taken into account the fact that $g/g_\rho\sim t_\theta$, and another one involving the bottom Yukawa, of order $\sim \epsilon^2\Delta_{Q^b}^2y_{b({\bf1},{\bf2})}^2/m_{Q^b}^2m_B^2$. Although the corrections mediated by $\Delta_{Q^b}$ are suppressed by a small mixing compared with those mediated by $\Delta_{Q^t}$, the large couplings of the {\it composite} sector can compensate that suppression, mainly with the term proportional to $t_{\theta}^{-2}$. Also notice that the contribution from mixing with $Q^t$ is negative, whereas the contribution from mixing with $Q^b$ is positive, as expected from the representation under SO(5) chosen for this fermion. The presence of a negative contribution to $\delta g_{b_L}^{\rm mass}$ requires a mixing $\Delta_{Q^b}$ larger than expected to compensate the negative term and obtain a positive $\delta g_{b_L}^{\rm mass}$. The correction to the Right-coupling in the mass basis is positive and is controlled by the mixing with $B$. It can also be enhanced by large {\it composite} couplings.

\section{Low-energy effective theory}\label{leff}
We consider in this section the effective theory at energies lower than the scale of resonances. This is particularly useful when studying the $Zb\bar b$-interactions, because one can gain understanding on the different symmetries that can protect the couplings, as well as on the origin of the sources that break those symmetries and induce corrections to the couplings. The effective theory also provides a very compact formalism to describe the spectrum and therefore to compute the Higgs potential at one-loop. We proceed by integrating-out the states of the {\it composite} sector, obtaining an effective theory for the {\it elementary} ones. 

The gauge sector has the same symmetries and field content as Ref.~\cite{Carena:2014ria}, where the effective Lagrangian for the gauge fields was computed in detail. Below we show the main results that are needed for the purposes of this article and in Ap.~\ref{App-correlators} we show the gauge correlators in the SO(4)-symmetric vacuum. The quadratic effective Lagrangian for the gauge fields arising from integrating-out the {\it composite} vectors is:
\begin{align}\label{Leff-gauge-sources}
{\cal L}_{\rm eff}\supset \frac{1}{2}\sum_r\Pi^A_rP_r(U^\dagger a_\mu)\ P_r(U^\dagger a^\mu)+\frac{1}{2}\Pi^Xx_\mu x^\mu \ ,
\end{align}
with $r$ being SO(4)-representations and $U$ acting on the {\bf 10} representation of SO(5). The correlators $\Pi^A_r$ can be computed straightforward by integrating out the {\it composite} vectors in the SO(4)-symmetric vacuum $\epsilon=0$. Although we are using a different parametrization of the pNGB field compared with Ref.~\cite{Carena:2014ria}, where it was parametrized in terms of a vector in the fundamental of SO(5), both descriptions coincide. By considering an arbitrary vacuum and keeping only the {\it elementary} gauge fields of SU(2)$_L\times$U(1)$_Y$, we obtain:
\begin{align}\label{Leff-gauge}
{\cal L}_{\rm eff}\supset \frac{1}{2} \, \sum^3_{i=1} \Pi_{w^i_L} \tilde{w}^i_{L \, \mu} \tilde{w}^{i \, \mu}_L
+ \Pi_{w^3_L \, b} \, \tilde{w}^3_{L \, \mu} \tilde{b}^{\mu}
+ \frac{1}{2} \, \Pi_{b} \, \tilde{b}_{\mu} \tilde{b}^{\mu}~,
\end{align}
where the correlators $\Pi_{w^i_L}$, $\Pi_{b}$ and $\Pi_{w^3_L \, b}$ can be expressed in terms of the correlators $\Pi_A^r$ and $\Pi_X$ as:
\begin{align}\label{correlators-gauge}
&\Pi_{w^i_L}= \Pi^A_{({\bf3},{\bf1})+({\bf1},{\bf3})} + \frac{1}{2} (\Pi^A_{({\bf2},{\bf2})} - \Pi^A_{({\bf3},{\bf1})+({\bf1},{\bf3})}) \sin^{2}\left(\frac{v}{f_h}\right) \, , \nonumber \\
&\Pi_{b}= \Pi^{X} + \Pi^A_{({\bf3},{\bf1})+({\bf1},{\bf3})} + \frac{1}{2} (\Pi^A_{({\bf2},{\bf2})} - \Pi^A_{({\bf3},{\bf1})+({\bf1},{\bf3})}) \sin^{2}\left(\frac{v}{f_h}\right) \, ,\nonumber \\
&\Pi_{w^3_L \, b}= - \frac{1}{2} (\Pi^A_{({\bf2},{\bf2})} - \Pi^A_{({\bf3},{\bf1})+({\bf1},{\bf3})}) \sin^{2}\left(\frac{v}{f_h}\right) \, .
\end{align}

At quadratic level in the {\it elementary} fermions, the most general effective Lagrangian arising from integrating-out the {\it composite} states of our two-site model can be written as:
\begin{align}\label{Leff-fermions-sources}
{\cal L}_{\rm eff}\supset
\sum_{\psi=q^t,q^b,t,b}\sum_{r}\overline{P_r(U^\dagger\psi)}\ \pslash\ \Pi^\psi_r\ P_r(U^\dagger\psi)
+ \sum_{\psi=t,b}\sum_{r}\overline{P_r(U^\dagger q^\psi)}\ M^\psi_r\ P_r(U^\dagger \psi) + {\rm h.c.}
\end{align}
with $U$ acting on the SO(5)-representation corresponding to each fermion. The correlators $\Pi^\psi_r$ and $M^\psi_r$ can be computed straightforward by considering the SO(4)-symmetric vacuum $\epsilon=0$. For a general vacuum, and keeping only the dynamical fields, Eq.~(\ref{Leff-fermions-sources}) leads to~\cite{Carena:2014ria}:
\begin{eqnarray}
\label{Leff-fermions}
{\cal L}_{\rm eff}&\supset& \bar t_L \pslash (Z_q+\Pi_{t_L}) t_L + \bar b_L \pslash (Z_q+\Pi_{b_L}) b_L + \bar t_R \pslash (Z_t+\Pi_{t_R}) t_R + \bar b_R \pslash (Z_b+\Pi_{b_R}) b_R \nonumber \\ [0.4em]
& & \mbox{} + \bar t_L M_t t_R + \bar b_L M_b b_R + {\rm h.c.}
\end{eqnarray}
where $Z_\psi$ are the kinetic terms of the {\it elementary} fermions. The correlators of the dynamical fermions can be expressed in terms of the ones in the SO(4)-symmetric vacuum as: 
\begin{align}
\Pi_{t_{L}} =& \alpha^{(2,2)}_{t_L,q^t}(h) \Pi^{q^t}_{(2,2)} + \alpha^{(1,1)}_{t_L,q^t}(h) \Pi^{q^t}_{(1,1)} \nonumber\\
&+ \alpha^{(2,1)}_{t_L,q^b}(h)\Pi^{q^b}_{(2,1)} + \alpha^{(1,2)}_{t_L,q^b}(h) \Pi^{q^b}_{(1,2)} +
\alpha^{(2,3)}_{t_L,q^b}(h) \Pi^{q^b}_{(2,3)} + \alpha^{(3,2)}_{t_L,q^b}(h)\Pi^{q^b}_{(3,2)} \ , \nonumber\\
\Pi_{b_{L}} =& \alpha^{(2,2)}_{b_L,q^t}(h) \Pi^{q^t}_{(2,2)} + \alpha^{(1,1)}_{b_L,q^t}(h) \Pi^{q^t}_{(1,1)} \ , \nonumber\\
&+ \alpha^{(2,1)}_{b_L,q^b}(h)\Pi^{q^b}_{(2,1)} + \alpha^{(1,2)}_{b_L,q^b}(h) \Pi^{q^b}_{(1,2)} + \alpha^{(2,3)}_{b_L,q^b}(h) \Pi^{q^b}_{(2,3)} + \alpha^{(3,2)}_{b_L,q^b}(h)\Pi^{q^b}_{(3,2)} \ , \nonumber \\
\Pi_{t_{R}} =& \alpha^{(2,2)}_{t_R,t}(h) \Pi^t_{(2,2)} + \alpha^{(1,1)}_{t_R,t}(h)\Pi^t_{(1,1)} \ , \nonumber\\
\Pi_{b_{R}} =& \alpha^{(2,1)}_{b_R,b}(h) \Pi^b_{(2,1)} + \alpha^{(1,2)}_{b_R,b}(h)\Pi^b_{(1,2)} \ , \nonumber \\
M_{t} =& \beta^{(2,2)}_{t}(h) M^t_{(2,2)} + \beta^{(1,1)}_{t}(h) M^t_{(1,1)} \ , \nonumber\\
M_{b} =& \beta^{(2,1)}_{b}(h) M^b_{(2,1)} + \beta^{(1,2)}_{b}(h) M^b_{(1,2)} \ ,
\end{align}
where the functions $\alpha$ and $\beta$ can be obtained by computing the invariants in an arbitrary vacuum. For those related with the top quark we obtain:
\begin{align}
&\alpha^{(2,2)}_{t_L,q^t}(h)=\frac{1}{4}(3+c_{2h}) \ ,
&\alpha^{(1,1)}_{t_L,q^t}(h)&=\frac{1}{2} s_{h}^{2} \ , 
\nonumber \\
&\alpha^{(2,1)}_{t_L,q^b}(h)=0 \ , 
&\alpha^{(1,2)}_{t_L,q^b}(h)&=0 \ ,
&\alpha^{(2,3)}_{t_L,q^b}(h)=c_{\frac{h}{2}}^{2} \ ,
\qquad \alpha^{(3,2)}_{t_L,q^b}(h)&=s_{\frac{h}{2}}^{2} \ , 
\nonumber \\
&\alpha^{(2,2)}_{t_R,t}(h)=s_{h}^{2} \ , 
&\alpha^{(1,1)}_{t_R,t}(h)&=c_{h}^{2} \ ,
\nonumber \\
&\beta^{(2,2)}_{t}(h)=\frac{c_{h} s_{h}}{\sqrt{2}} \ , 
&\beta^{(1,1)}_{t}(h)&=-\frac{c_{h} s_{h}}{\sqrt{2}} \ ,
\nonumber
\end{align}
whereas for those related with the bottom-quark:
\begin{align}
\alpha^{(2,2)}_{b_L,q^t}(h) &= 1 \ , & \alpha^{(1,1)}_{b_L,q^t}(h) &= 0 \ , \nonumber \\
\alpha^{(2,1)}_{b_L,q^b}(h) &= \frac{5}{8} s_{\frac{h}{2}}^{2} s_{h}^{2} \ ,
& \alpha^{(1,2)}_{b_L,q^b}(h) &= \frac{5}{128} \frac{4 c_{2h} - c_{4h} - 3}{c_{h} - 1} \ , \nonumber \\
\alpha^{(2,3)}_{b_L,q^b}(h) &= c_{\frac{h}{2}}^{2} \frac{3 c_{2h} + 4c_{h} + 9}{16} \ , 
&\alpha^{(3,2)}_{b_L,q^b}(h) &= s_{\frac{h}{2}}^{2}\frac{3 c_{2h} -4 c_{h} + 9}{16} \nonumber \ ,\\
\alpha^{(2,1)}_{b_R,b}(h) &= s_{\frac{h}{2}}^{2} \ , 
&\alpha^{(1,2)}_{b_R,b}(h) &=c_{\frac{h}{2}}^{2} \ ,\nonumber \\
\beta^{(2,1)}_{b}(h) &= \frac{i}{2} \sqrt{\frac{5}{2}}s^{2}_{\frac{h}{2}} s_{h} \ , 
&\beta^{(1,2)}_{b}(h) &=-\frac{1}{4} \sqrt{\frac{5}{2}} (1 + c_{h}) s_{h} \ .
\end{align}
In the previous expressions we have used the following shorthand notation $c_{nh}\equiv \cos\left(n\frac{h}{f_h}\right)$ and similarly for $s_{nh}$.
The correlators in the SO(4)-symmetric vacuum expressed in terms of the parameters of the model are shown in Ap.~\ref{App-correlators}.

Following Ref.~\cite{Barbieri:2004qk} we define:
\begin{align}\label{matching-eff-th}
&v_{SM}^2=\Pi_{w^1_L}(0)=f_h^2\epsilon^2 \ , \nonumber
\\
&\frac{1}{g^2}=\frac{1}{g_0^{2}}+\Pi_{w^1_L}'(0)=\frac{1}{g_0^2}+\frac{1}{g_\rho^2}-\frac{\epsilon^2}{g_\rho^2}\frac{f_1^2(2f_\Omega^2+f_1^2)}{(f_\Omega^2+f_1^2)^2} \ , \nonumber
\\
&\frac{1}{{g'}^2}=\frac{1}{g_0^{'2}}+\Pi_{b}'(0)=\frac{1}{g_0^{'2}}+\frac{1}{g_\rho^2}+\frac{1}{g_X^2}-\frac{\epsilon^2}{g_\rho^2}\frac{f_1^2(2f_\Omega^2+f_1^2)}{(f_\Omega^2+f_1^2)^2} \ ,
\end{align}
where $\Pi(0)\equiv\Pi(p^2)_{p^2=0}$. The matching of Eq.~(\ref{matching-eff-th}) implies that in the effective theory, at zero momentum the corrections to the gauge interactions are mediated by the mixing with the heavy fermions, such that, for zero fermionic mixing the coupling is SM-like. We will show the explicit results for the $Z$ interactions in the effective theory in sec.~\ref{Zcouplingseffth}.

\subsection{$Z$-interactions in the low energy effective theory}\label{Zcouplingseffth}
In this section we compute the $Z$-interactions in the effective theory. Since the {\it elementary} and {\it composite} fields have well defined transformation properties under the gauge symmetry groups and the effective theory is formulated in terms of the {\it elementary} fields after integrating-out the {\it composite} ones, the symmetries of the model are manifest in the low energy effective theory. In particular, we expect the symmetries protecting several couplings to manifest explicitly in this basis; we will show below that this is the case. We will consider the couplings at $p^2=0$.

We begin by analysing the interactions with the top quark. Since $t_R$ transforms as $({\bf 1},{\bf 1})$, from $P_C$ symmetry~\cite{Agashe:2006at} we expect the $t_R$ coupling to be protected. We have checked that property by explicit calculation, finding no corrections. On the other hand the $t_L$ coupling is not protected. In Eq.~(\ref{eqZcouplingseff}) we show our result. Although it is possible to obtain the coupling to all orders in $\epsilon$, for simplicity in the presentation we only show the leading order terms.

The $b_L$ coupling can receive contributions from the mixing with $Q^t$ and $Q^b$, the first potentially large due to the large $\Delta_{Q^t}$ required by the top mass. However, since the composite fermion in $Q^t$ mixing with $b_L$ has $T^{3L}=T^{3R}=-1/2$, a $P_{LR}$ symmetry protects this coupling from corrections induced by mixing with $Q^t$~\cite{Agashe:2006at}. On the other hand, the mixing with $Q^b$ induces a positive shift of the $b_L$ coupling, suppressed by the small $\Delta_{Q^b}$ mixing that controls the bottom mass~\cite{DaRold:2010as}. Notice that it is positive and is suppressed by $\epsilon^2\Delta_{Q^b}^2/m_{Q^b}^2$.

The shift of the $b_R$ coupling is positive and controlled by $\Delta_B^2$. 

To leading order in $\epsilon$ we obtain:
\begin{align}\label{eqZcouplingseff}
\delta g^{\rm eff}_{t_L}=\frac{g}{c_W}\frac{\epsilon^2}{4}& \Big\{-\Delta_{Q^t}^2m_{Q^b}^2\ \left[f_\Omega^2\left(y_{t({\bf 1},{\bf 1})}-y_{t({\bf 2},{\bf 2})}\right)^2+f_1^2\left(2m_T^2+y_{t({\bf 1},{\bf 1})}^2-4y_{t({\bf 1},{\bf 1})}y_{t({\bf 2},{\bf 2})}+5y_{t({\bf 2},{\bf 2})}^2\right)\right]\nonumber\\
&\ \ +\Delta_{Q^b}^2f_1^2m_{Q^t}^2m_T^2\Big\}\nonumber\\
&\left\{(f_\Omega^2+f_1^2)\left[m_{Q^t}^2m_T^2(2m_{Q^b}^2+\Delta_{Q^b}^2)+m_{Q^b}^2\Delta_{Q^t}^2(m_T^2+y_{t({\bf 2},{\bf 2})}^2)\right]\right\}^{-1}  +{\cal O}(\epsilon^4)\ ,
\nonumber \\
\delta g^{\rm eff}_{t_R}=0 \ , \ \ \ \ &
\nonumber \\
\delta g^{\rm eff}_{b_L}=\frac{g}{c_W}\frac{\epsilon^2}{16}& \Delta_{Q^b}^2\ \frac{m_{Q^t}^2m_{T}^2\left[12f_1^2 m_{B}^2+5y_{b({\bf 1},{\bf 2})}^2\left(f_\Omega^2+f_1^2\right)\right]}{m_B^2(f_\Omega^2+f_1^2)\left[m_{Q^t}^2m_T^2\left(2m_{Q^b}^2+\Delta_{Q^b}^2\right)+m_{Q^b}^2\left(m_T^2+y_{t({\bf 2},{\bf 2})}^2\right)\right]} +{\cal O}(\epsilon^4)\ ,
\nonumber \\
\delta g^{\rm eff}_{b_R}=\frac{g}{c_W}\frac{\epsilon^2}{8}& \Delta_B^2\ \frac{f_\Omega^2\left(y_{b({\bf 1},{\bf 2})}^2+y_{b({\bf 2},{\bf 1})}^2\right)+f_1^2\left(2m_{Q^b}^2+5y_{b({\bf 1},{\bf 2})}^2+y_{b({\bf 2},{\bf 1})}^2\right)}{(f_\Omega^2+f_1^2)\left[m_B^2m_{Q^b}^2+\left(m_{Q^b}^2+y_{b({\bf 1},{\bf 2})}^2\right)\Delta_{Q^b}^2\right]}  +{\cal O}(\epsilon^4)\ .
\end{align}
The Weinberg angle has been defined by the usual tree-level relation, with the couplings $g$ and $g'$ defined in the effective theory by Eq.~(\ref{matching-eff-th}).

By comparing these effective couplings with the couplings between the mass eigenstates one can appreciate the size of the corrections, the most important one being the contribution to $\delta g_{b_L}$ arising from mixing with $Q^t$ that is present in the basis of mass eigenstates, but is not present in the effective Lagrangian thanks to the $P_{LR}$-symmetry.

For non zero momentum the $Z$-interactions become form factors with non trivial momentum dependence. These corrections also give rise to new Lorentz structures~\cite{Agashe:2005dk}, as: $p_\mu\pslash$, $p'_\mu\pslash$, $p_\mu\pslash'$, $p'_\mu\pslash'$, $\gamma_\mu\pslash$, $\gamma_\mu\pslash'$, with $p_\mu$ and $p'_\mu$ the momentum of the particle and antiparticle. Note that the last two structures flip chiralities.

\section{Higgs potential}\label{higgspotential}
Two site models lead to a one-loop Higgs potential that is finite and calculable, provided that one excludes certain chiral structures in the Yukawa interactions~\cite{DeCurtis:2011yx,Carena:2014ria}, as detailed in sec.~\ref{model}.~\footnote{Another possibility would be to allow for all the possible chiral structures in a model with at least three sites~\cite{Panico:2011pw}.}

We assume that the light generations have small mixing for both chiralities and therefore do not have a large impact in the one-loop Higgs potential. The effect of the other states are fully taken into account. Using the correlators of the effective theory, the Higgs potential at one loop can be written as~\cite{Agashe:2004rs,Carena:2014ria}:
\begin{align}\label{vh}
V(h)=\int \frac{d^4p}{(2\pi)^4} &\left\{3\sum^2_{i = 1} \log \left(\frac{p^2}{g_0^2}+\Pi_{w^i_L}\right) + \frac{3}{2}\log \left[\left(\frac{p^2}{g_0^2}+\Pi_{w^3_L}\right) \left(\frac{p^2}{g_0^{'2}}+\Pi_{b}\right) - \Pi_{w^3_L \, b}^{\ 2}\right]
\right.
\nonumber \\
&\hspace{1cm}\left.- 2N_c\sum_{\psi=t,b} \log\left[p^2(Z_{\psi_L}+\Pi_{\psi_L})(Z_{\psi_R}+\Pi_{\psi_R}) - |M_{\psi}|^2\right] \right\} \ .
\end{align}
Notice that we have included the kinetic terms in the expression for the potential because they were not included in the definitions of the correlators.

The fermionic correlators are proportional to the mixing squared, $\Pi_{L,R}\propto\Delta_{L,R}^2$ and $M\propto\Delta_L\Delta_R$. Thus we expect fermions with large mixing to dominate the potential.

The top quark requires large mixing with the {\it composite} sector to account for its large mass. The mixing explicitly breaks the symmetry behind the NGB nature of the Higgs, thus the top quark usually dominates the Higgs potential. A large correction to $Zb_R\bar b_R$ also requires a considerable mixing of $b_R$, with a potentially important effect in $V(h)$. However, for $y_{b({\bf 1},{\bf 2})}=y_{b({\bf 2},{\bf 1})}$ as in the present model, the correlator $\Pi_{b_R}$ is independent of $h$. The reason is that in this limit the only invariant involving just $b_R$ is the trivial one. As we will discuss in the next sections, the mixing of $b_L$ is smaller than those of the top and $b_R$, therefore we expect the effect of the bottom quark in the Higgs potential to be subleading in the present model.

For $h=0$ there is a divergent contribution to the Higgs potential that has no impact on EWSB. A finite and meaningful potential can be obtained by computing $V(h)-V(0)$. 

\section{Numerical results}\label{Zcouplingsnumeric}
One of the goals of this article is to obtain the shift of $Zb\bar b$ couplings in the region of the parameter space of the model where the EW symmetry is broken and the masses of the SM fields are around their physical values. To satisfy these conditions we have performed a random scan over the parameters of the model, selecting the proper points. Below we describe the implementation of the random scan and after that we present our results for the couplings. We have followed a similar procedure to the one of Ref.~\cite{Carena:2014ria}.

We have scanned over the bosonic mixing $t_\theta$ and $t_{\theta_X}\equiv \tan\theta_X=g_x/g_X$. For simplicity we have imposed the relation $t_\theta=t_{\theta_X}$, that fixes the ratio between {\it elementary} and {\it composite} couplings to be the same for the different groups. We have fixed the remaining freedom in the set of gauge couplings by matching with the SM couplings, as shown in Eq.~(\ref{matching-massless}). We have expressed the decay constants of the $\sigma$-model fields in terms of the masses of the vectors $m_{\tilde\rho}$ and the Higgs decay constant $f_h$, and we have scanned over them. The choice $t_\theta=t_{\theta_X}$ leads to $g_X\sim 0.65 g_\rho$, thus to avoid a light vector resonance arising from the U(1)$_X$ symmetry we have considered $f_{\Omega_X}\sim 2f_{\Omega}$.

For the fermions we have scanned over the mixing $t_{\theta_\Psi}$, with $\Psi=Q^t,Q^b,T$ and $B$. We have also expressed the {\it composite} masses $m_\Psi$ in terms of the masses $m_{\tilde\Psi}$ (see the discussion in the paragraphs below Eq.~(\ref{matching-massless})). For simplicity we have fixed $m_{\tilde\Psi}=m_{\tilde\rho}$, the same value for all the representations, however notice that even under this simplifying assumption there can be large splittings between the different fermions due to the suppression of the masses of the custodians when the mixing is large. Besides we have also scanned over the Yukawa couplings of the composite sector: $y_{t({\bf1},{\bf1})},y_{t({\bf2},{\bf2})}$ and $y_{b({\bf1},{\bf2})}=y_{b({\bf2},{\bf1})}$.

Concerning the numerical values, we have allowed large ranges for all the parameters. However we present results constrained to some regions of the parameter space, motivated by phenomenological considerations. We let the bosonic mixing vary according to: $t_\theta\in[0.11,0.33]$, we have allowed small values for this parameter because they favour larger shifts in the $Z$ couplings. We have considered $s_{\theta_{Q^t}},s_{\theta_T}\in[0.5,1)$ for the top, $s_{\theta_{Q^b}}\in[0.05,0.45]$ and $s_{\theta_B}\in[0.1,0.9]$. For $f_h$ we scanned over $f_h\in[0.5,2.8]$~TeV, the smallest value corresponding to $\epsilon\sim 0.5$ for $v_{SM}=246$~GeV, of the order of the maximum value allowed by EWPT. For the masses of the resonances we have allowed $m_{\tilde\rho}\sim2-8$~TeV, that is consistent with the approximate relation $m_\rho\sim g_\rho f_h$, with $g_\rho$ in the range determined by the bosonic mixing angle detailed before. We have considered real Yukawa couplings within the NDA perturbative regime: $|y/f_h|\leq2\pi$.

For each point we have evaluated the one-loop Higgs potential, as well as the masses of the Higgs, the $Z$ boson, the top and bottom. We have normalized all the dimensional parameters by demanding the mass of the lightest neutral boson to reproduce the SM $Z$ mass, and we have only kept the points that match the SM values for the other masses. The phenomenology that we have studied is not very sensitive to small variations around the central values of the masses, therefore, due to the finite time for CPU calculation, we have considered the following ranges: $110\lesssim m_h\lesssim 140$~GeV, $130\lesssim m_t\lesssim 170$~GeV and $1\lesssim m_b\lesssim 4$~GeV. We have also discarded points with $\epsilon>0.5$ as well as points with masses of bosonic vector resonances below 2 TeV. After selecting the points that satisfy these conditions, we have computed the couplings by performing a numerical diagonalization. We have checked our analytic formulas by comparing with the results obtained by the procedure already described.

\begin{figure}
\centering
\psfrag{0.005}{$0.005$}
\psfrag{0.010}{$0.010$}
\psfrag{0.015}{$0.015$}
\psfrag{0.020}{$0.020$}
\psfrag{-0.002}{$-0.002$}
\psfrag{0.000}{$0$}
\psfrag{0.002}{$0.002$}
\psfrag{0.004}{$0.004$}
\psfrag{0.006}{$0.006$}
\psfrag{dL}{$\delta g_{b_L}$}
\psfrag{dR}{$\delta g_{b_R}$}
\includegraphics[width=\textwidth]{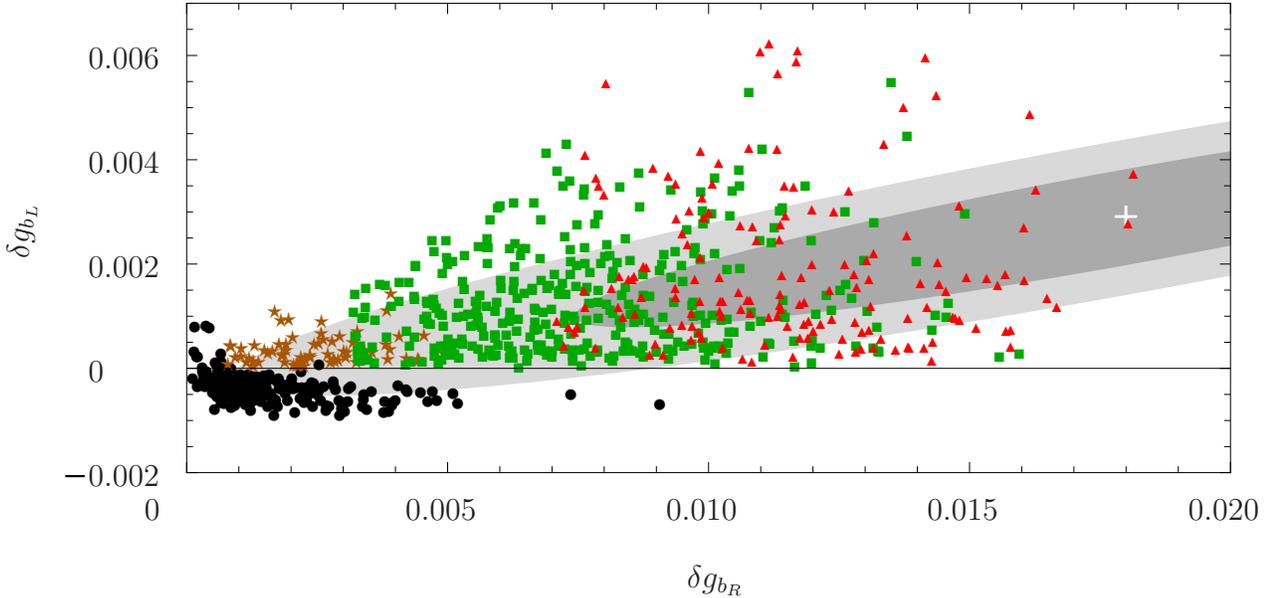}
\caption{Results for the shifts in the $Zb\bar b$ couplings. The ellipses correspond to the 68\% (dark region) and 95\% (light region) probability distribution for $\delta g_{b_L}$ and $\delta g_{b_R}$ from Ref.~\cite{Ciuchini:2014dea}, the white cross shows the center of the ellipses. The black points (dots) have bottom Yukawa coupling: $y_b/f_h\sim g_\rho$ and one small bottom mixing, typically the Left-handed one. The colored points (stars, squares and triangles) have sizeable $s_{\theta_{Q^b}}$ and $s_{\theta_B}$, and small bottom Yukawa coupling: $y_b/f_h\sim{\cal O}(10^{-1})$. The colors codify the value of $\epsilon$: orange points (stars) for $\epsilon\in(0,0.2)$, green points (squares) for $\epsilon\in(0.2,0.35)$ and red points (triangles) for $\epsilon\in(0.35,0.5)$.}
\label{figdeltag}
\end{figure}

In Fig.~\ref{figdeltag} we show $\delta g_{b_L}$ and $\delta g_{b_R}$ for the points of the scan that pass the selection conditions detailed in the previous paragraph, as well as the 68\% and 95\% probability distributions for these shifts from Ref.~\cite{Ciuchini:2014dea}. The white cross shows the center of the ellipses. First, one can see that there are many points that can solve the discrepancy attributed to the deviation in $A^b_{FB}$. Second, the probability distributions of Ref.~\cite{Ciuchini:2014dea} present a strong correlation between both shifts, such that shifting just one of the couplings does not improve the fit considerably. Instead the probability distributions prefer a shift $\delta g_{b_L}\sim{\cal O}(10^{-3})$ correlated with a shift $\delta g_{b_R}\sim{\cal O}(10^{-2})$. In order to shift both couplings simultaneously, considerable mixing $s_{\theta_{Q^b}}$ and $s_{\theta_B}$ are required, as shown in Eqs.~(\ref{eqZcouplingsmass}). In that case, in order to reproduce the bottom mass, since the same mixing that controls the shift of the couplings also controls the mass of the bottom quark, the {\it composite} bottom Yukawa coupling $y_{b({\bf1},{\bf2})}$ has to be suppressed to compensate the effect of the mixing and lead to a small bottom mass. On the other hand, if the {\it composite} bottom Yukawa coupling is large, of the same size as the other {\it composite} couplings, at least the mixing of one of the chiralities has to be small to suppress the mass, leading to a very small shift for the coupling associated to that chirality. This behaviour is present independently of the nature of the composite Higgs, whether it is a pNGB or not and is instead associated to partial compositeness, and has been noticed before in scenarios where the Higgs was not a pNGB~\cite{DaRold:2010as}. That behaviour is shown in Fig.~\ref{figdeltag}, where the black points (dots) correspond to regions of the parameter space where the Yukawa is large: $y_b/f_h\sim g_\rho$, whereas colored points (stars, squares and triangles) correspond to regions of the parameter space with sizeable $s_{\theta_{Q^b}}$ and $s_{\theta_B}$, and small bottom Yukawa coupling: $y_b/f_h\sim{\cal O}(10^{-1})$, introducing a little hierarchy between the {\it composite} couplings. This Yukawa violates the assumption that $g_{\rho}\gg g_{SM}$ for all the {\it composite} couplings. The colors codify the value of $\epsilon$: orange points (stars) for $\epsilon\in(0,0.2)$, green points (squares) for $\epsilon\in(0.2,0.35)$ and red points (triangles) for $\epsilon\in(0.35,0.5)$. As expected, since to leading order in $\epsilon$ the shift is proportional to $\epsilon^2$, a large shift requires $\epsilon\gtrsim 0.2$. We have not found points that satisfy all the selection criteria specified before and produce corrections that lie outside of the range shown in the plot. 

The previous results introduce at least two different scales: one for the Yukawa of the top sector and another one for the Yukawa of the bottom sector. Still, since we have considered only the third generation of quarks, the present analysis does not give any information about the flavor structure of the Yukawa couplings and composite masses. That issue is beyond the scope of this work.

\section{Radiative corrections to $Zb\bar b$}\label{Zcouplingsloop}
We have shown the tree level corrections to $Zb\bar b$ interactions, however it is important to estimate the size of the one-loop contributions, to test the stability of the corrections. In particular, the large mixing of some of the fermions can have an impact on the small correction allowed for $Zb_L\bar b_L$. Ref.~\cite{Anastasiou:2009rv} has considered the one-loop correction to this interaction in models with SO(5)/SO(4) symmetry breaking pattern, computing the contributions from fermions in the representation ${\bf 5}_{\bf 2/3}$.  The authors argued that when there are several multiplets present in the theory, in large regions of the parameter space EWPT can be satisfied, particularly a shift of $g_{b_L}$ small enough is obtained. There are some simplifications in the case ${\bf 5}_{\bf 2/3}$ because there is no $b_R$ partner in the composite sector and also because the heavy down-type quarks, the quarks with charge $Q=-1/3$, arise from multiplets with the same quantum numbers under SU(2)$_L\times$U(1)$_Y$ as the SM doublet: ${\bf 2}_{\bf 1/6}$. Since in our model there are quarks with exotic hypercharge assignments as well as large mixing for $b_R$, it is worth performing an estimation of the one-loop corrections in the present case.

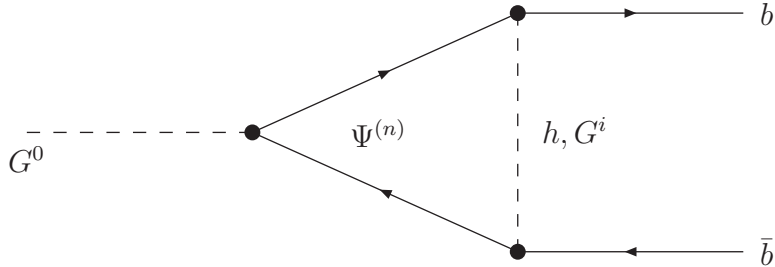
\begin{figure}
\centering
\begin{picture}(300,110)
        \DashLine(0,50)(85,50){5}
	\ArrowLine(85,50)(185,95)
	\ArrowLine(185,5)(85,50)
	\Vertex(85,50){3}
	\Vertex(185,95){3}
	\Vertex(185,5){3}
        \DashLine(185,95)(185,5){5}
	\ArrowLine(185,95)(270,95)
	\ArrowLine(270,5)(185,5)
        \Text(0,40)[]{$G^0$}
        \Text(133,50)[]{$\Psi^{(n)}$}
        \Text(195,50)[l]{$h,G^i$}
        \Text(280,95)[]{$b$}
        \Text(280,5)[]{$\bar b$}
\end{picture}
\caption{Radiative correction to $Zb\bar b$ at 1-loop in the gaugeless limit, $G^{0/i}$ is the neutral/charged NGB associated with the longitudinal degree of freedom of the $Z/W^i$ and $\Psi^{(n)}$ is a heavy fermionic resonance.}
\label{loopdiag}
\end{figure}

We take the gaugeless limit for our calculation, which has been shown before to yield good predictions while simplifying the calculations greatly \cite{Barbieri:1992nz}. The $Zb\overline{b}$ coupling will thus be estimated by calculating the $G^{0}b\overline{b}$ coupling, where $G^{0}$ is the NGB eaten by the $Z$. By taking this limit, we are left with the diagram shown in Fig.~\ref{loopdiag}. This diagram can be split into various contributions corresponding to corrections to $G^{0}b_{L}\overline{b_{L}}$ and $G^{0}b_{R}\overline{b_{R}}$ and to fermions $\Psi^{(n)}$ of charges $-4/3$ (v-type quarks), $-1/3$ (down-type quarks) or $2/3$ (up-type quarks) running in the loop.~\footnote{With the NGB in the loop changing accordingly to maintain conservation of electric charge in the vertices.} This diagram represents a finite contribution to the $G^{0}b\overline{b}$ coupling which can be explicitly calculated for all fermions running in the loop. Details on the expressions obtained for this diagram can be found for example in \cite{Oliver:2002up,Chivukula:2010nw}. 

Performing a numerical calculation for different sets of points in the parameter space of our model we arrive at the following results. We find that the shift to the $G^{0}b_{L}\overline{b_{L}}$ coupling produced by considering only the charge $2/3$ fermions running in the loop, subtracting the SM contribution, is of order $\delta g_{b_L}|_{T}\sim 10^{-4}$. Similarly, $\delta g_{b_R}|_{T}$ is of order $\sim 10^{-5}$. The difference in size for these shifts can be readily understood by considering a perturbative expansion in both the fermionic mixing and the Yukawa couplings with the NGB's. For $\delta g_{b_L}|_{T}$, we would expect the main perturbative contribution to be proportional to $s_{\theta_{Q^t}}^{2} y_{t}^{4} L_{T}$, where $L_{q}$ is the loop factor coming from the phase space integration with the q-type quarks running in the loop. This is due to the fact that the elementary $b_{L}$ has a non-zero projection on an elementary ${\bf5}$ which, after mixing with the composite ${\bf5}$, can couple directly to a top-like fermion through the Yukawa coupling with the charged NGB's. On the other hand, the $b_{R}$ field is embedded in a ${\bf4}$ which contains no charge $2/3$ fermions and which must undergo various mixing and Yukawa couplings before being able to mix with the top-like fermions. In particular, it mixes with the elementary $q_{L}$ so it can then follow a path similar to the one explained for the $b_{L}$. Hence, we expect the largest contribution to $\delta g_{b_R}|_{T}$ to be $\sim s_{\theta_B}^{2} y_{b}^{2} s_{\theta_{Q^b}}^{2} y_{t}^{2} L_{T}$. In order to get the right mass for the SM $b$ quark the product $(s_{\theta_{Q^b}}s_{\theta_B}y_b)$ needs to be small, thus $\delta g_{b_R}|_{T}$ is suppressed with respect to $\delta g_{b_L}|_{T}$. A large tree level shift to the $Zb_{R}\overline{b_{R}}$ coupling typically requires $s_{\theta_B}$ large, then in this case the suppression in $\delta g_{b_R}|_{T}/\delta g_{b_L}|_{T}$ is of order $s_{\theta_{Q^b}}^{2}y_b^2/y_t^2$.

Similarly, we can compare the shifts produced by the top-like quarks with those produced by the charge $(-4/3)$ quarks. For the v-type quarks, we find $\delta g_{b_L}|_{V} \sim 10^{-8}$. This is $4$ orders of magnitude smaller that $\delta g_{b_L}|_{T}$, which is consistent with the fact the the main perturbative contribution to it is expected to be $\sim s_{\theta_{Q^b}}^{2} y_{b}^{4} L_{V}$ and thus heavily suppressed with respect to $\delta g_{b_L}|_{T}$. On the other hand, the main perturbative contribution to $\delta g_{b_R}|_{V} \sim s_{\theta_B}^{2} y_{b}^{4} L_{V}$, resulting in $\delta g_{b_R}|_{V} \sim 10^{-5}$ which is of the same size as $\delta g_{b_R}|_{T}$; in this case, the difference in loop factors, Yukawa couplings and fermionic mixing roughly compensates.

There is also a contribution due to bottom-type quarks running in the loop (together with a neutral NGB), which can be analyzed in a similar way, and whose size is negligible compared to that of the up-type quarks contribution.

It is worth mentioning that due to the nonlinear coupling of the NGB's in our model and the mixing of b-type quarks, other Feynman diagrams can be constructed. Some examples are diagrams modifying the self energy of the bottom~\cite{Abe:2009ni} and diagrams arising from 4-fermion operators~\cite{Grojean:2013qca}. In some cases these diagrams can be divergent and must be renormalized.

\section{Phenomenology at colliders}\label{pheno}
We briefly discuss the phenomenology of the model at colliders as the LHC. Similar to other realizations of a Higgs as pNGB of SO(5)/SO(4), we expect light fermionic resonances also known as custodians with charges $Q=+5/3,2/3$ and $-1/3$; these are usually present in models based on representations ${\bf 5}_{\bf 2/3}$, ${\bf 10}_{\bf 2/3}$ and ${\bf 14}_{\bf 2/3}$. In the present model, the representation ${\bf 4}_{-5/6}$ leads also to resonances with new exotic charges $Q=-4/3$, whereas the ${\bf 16}_{-5/6}$ leads to resonances with $Q=-4/3$ and $-7/3$. The masses of the former can have considerable suppression compared to $m_\rho$, due to the non-negligible mixing $s_{\theta_B}\sim0.6$, whereas for the later they are expected in the same range as $m_\rho$, since the suppression of their masses is small as long as $s_{Q^b}$ is not so large. Moreover, for small {\it composite} Yukawa in the bottom sector, the mirror fermion in ${\bf 4}_{\bf -5/6}$ contains three rather light custodians that are almost degenerate and, under SU(2)$_L\times$U(1)$_Y$, have the following quantum numbers: one doublet ${\bf 2}_{\bf -5/6}$ and one singlet ${\bf 1}_{\bf -4/3}$. Similarly, in this limit the ${\bf 16}_{-5/6}$ contains a set of almost degenerate custodians generically heavier than the previous ones, that under SU(2)$_L\times$U(1)$_Y$ can be organized as: ${\bf 3}_{\bf -4/3}$, ${\bf 3}_{\bf -1/3}$, ${\bf 2}_{\bf -11/6}$, two ${\bf 2}_{\bf -5/6}$ doublets, ${\bf 2}_{\bf 1/6}$, ${\bf 1}_{\bf -4/3}$ and ${\bf 1}_{\bf -1/3}$. For the points shown in Fig.~\ref{figdeltag} we find that generically the $b_R$ custodians have masses similar to the custodians of the top. The production and detection of exotic resonances associated to $b_R$ would be a distinctive signature of this kind of models. Refs.~\cite{Kumar:2010vx,Alvarez:2013qwa} have explored this scenario, selecting the single production of states with $Q=-4/3$ and proposing a search strategy at LHC, see also~\cite{AguilarSaavedra:2009es} for a very comprehensive analysis. The present model provides a complete framework to study the phenomenology of these exotic resonances given the presence of a light Higgs.

There can also be interesting signals associated to the neutral interactions between the SM $b$ quark, a down-type fermionic resonance $b^{(m)}$ and a neutral vector boson, either the SM $Z$ or a resonance $Z^{(n)}$, with $m=1, \dots 10$ and $n=1, \dots 5$, the indexes ordered by increasing mass. The different processes to create one of this resonances and its dominating decay channel depend on the spectrum and the size of the couplings. Although there are many different possibilities and the spectrum and couplings can change much as the parameters of the theory vary, we want to discuss some general situations that can be expected in the present type of scenarios. After that we will describe the generic properties of the spectrum and couplings that can lead to these processes. First, if the coupling $Zb^{(m)}\bar b$ is large, $b^{(m)}$ can be produced (either QCD production $b^{(m)}\bar b^{(m)}$ or single EW production) and decay to $Zb$, leading to a final state where one could fully reconstruct the $b^{(m)}$-mass by selecting a visible $Z$-decay channel. In the case of $Z^{(n)}$ heavier than $b^{(m)}$, if $Z^{(n)}$ is produced it can decay via $Z^{(n)}\to b^{(m)}\bar b$, and eventually to $Zb\bar b$ if the coupling $Zb^{(m)}\bar b$ is large enough. In this case the full process would be $\psi\bar\psi\to Z^{(n)}\to b^{(m)}\bar b\to Zb\bar b$, with bottom quarks and $Z$ of large $p_T$ and with the eventual possibility to reconstruct the full mass of $Z^{(n)}$ and $b^{(m)}$. In the case of production of $b^{(m)}$ heavier than $Z^{(n)}$, a very interesting decay would be a cascade of neutral decays like $b^{(m)}\to Z^{(n)}b\to b^{(\ell)}\bar bb\to Zb\bar b b$, again with final states with large $p_T$.

Let us now briefly discuss the spectrum of down-type fermionic resonances and neutral vector bosons. We expect two light custodians: one arising from the multiplet ${\bf 5}_{2/3}$ mixing with $t_R$ and another one from the multiplet ${\bf 4}_{-5/6}$ mixing with $b_R$. There are also five custodians from the ${\bf 16}_{-5/6}$, usually with larger masses than the previous ones since the $b_L$ mixing is smaller than the other mixing, as well as three more states with no mass suppression: one from the ${\bf 5}_{2/3}$ mixing with $q_L$, one from ${\bf 4}_{-5/6}$ and another from the ${\bf 16}_{-5/6}$. In our scan, after taking into account EWSB effects, we have found a splitting between two sets of resonances: two light resonances mostly given by the first two custodians described in the beginning of this paragraph, and eight heavier resonances. For the neutral vectors we find three states with similar masses mostly arising from the neutral components of SO(4) and from U(1)$_X$, and two heavier states mostly from the coset SO(5)/SO(4). We also find that the heavier set of fermions have masses similar to the lighter set of bosons.~\footnote{It should be noted, however, that this mass ordering is only a consequence of the fact that we have assumed the mass scale of the gauge bosons and all fermions to be the same; this could be easily altered by introducing different scales for the resonances.} 

We consider the strength of the coupling $Z^{(n)} b^{(m)}_{L/R} \overline{b}_{L/R}$; we call these couplings $g^{nm}_{b_{L/R}}$. In what follows, we will always refer to the size of the couplings in units of $\frac{g}{c_{W}}$. We use the points with low {\it composite} Yukawa coupling presented in Fig. \ref{figdeltag} as a testing ground. We analyse first the Right-handed couplings of $Z^{(1)}$. We find that typically $g^{19}_{b_R}\sim\mathcal{O}(1)$, with $b^{(9)}$ usually slightly heavier than $Z^{(1)}$, although the mass difference is small in this case and the ordering can be changed easily. Another relevant $Z^{(1)}$ coupling is that of $g^{13}_{b_R}$, which can be as large as $0.8$ but can also fluctuate all the way down to negligible strength, depending on the value of $s_{\theta_{Q^b}}$. Contrary to $b^{(9)}$, $b^{(3)}$ is generally lighter than $Z^{(1)}$, but again the splitting is small.

$Z^{(3)}$ behaves in the same way as $Z^{(1)}$ when considering Right-handed couplings. $Z^{(4)}$ and $Z^{(5)}$ both present couplings of strength $\sim 1.4$ with $b^{(1)}_{R} \overline{b}_{R}$ and $b^{(2)}_{R} \overline{b}_{R}$ \footnote{What actually seems to happen here is that there is only one b-type ``light'' quark which has a large coupling, but depending on the values of the parameters considered it will be either the lightest or the second lightest down-resonance. The same appears to be true for these two  when it comes to Left-handed couplings.\label{fn2lightestbm}}. No other Right-handed coupling is as relevant as these for the $Z$ resonances.

For the Left-handed chirality there is not a $Z^{(1)}$ coupling that is consistently large, but all of $b^{(1)}, b^{(2)}, b^{(3)}$ and $b^{(8)}$ show couplings of $\mathcal{O}(1)$ for a large portion of the points we studied. $b^{(8)}$ has approximately the same mass as $Z^{(1)}$ for our region of parameter space. $Z^{(2)}$ and $Z^{(3)}$ show the same qualitative behavior as $Z^{(1)}$ when it comes to couplings with $b^{(1)}_{L} \overline{b}_{L}$ and $b^{(2)}_{L} \overline{b}_{L}$. $Z^{(2)}$, however, also shows a consistently large coupling with $b^{(10)}_{L} \overline{b}_{L}$ (we find it of $\mathcal{O}(1)$ for all the points considered). For our points, the mass of  $b^{(10)}$ is only $\sim 1.05 - 1.25$ times that of $Z^{(2)}$. The Left-handed couplings for the heavier $Z^{(n)}$ resonances are not, in general, as relevant as these.

Lastly, we can consider the $Z b^{(m)}_{L/R} \overline{b}_{L/R}$ couplings. Out of these, none of the Left-handed couplings are relevant. For the Right-handed  chirality, on the other hand, the two lightest $b^{(m)}$ resonances can exhibit a coupling half as large as that of the SM $b_R$ quark (see footnote \ref{fn2lightestbm}).

\section{Conclusions}\label{conclusions}
We have presented a model with a naturally light Higgs, arising as a pNGB of an SCFT, and able to induce a shift in the $Zb\bar b$ couplings that can relax the tension in $A_{FB}^b$ measured at LEP/SLD. The model is based on the breaking SO(5)/SO(4) and on the presence of {\it composite} fermions associated to the bottom sector transforming with the representations ${\bf4}$ and ${\bf16}$ of SO(5). These representations allow to generate the appropriate pull in the $Zb\bar b$ couplings and simultaneously to produce the small bottom mass. The top partners are embedded in a ${\bf5}$, such that no large corrections are induced for $Zb_L\bar b_L$ in the presence of large mixing for the top, these mixing trigger EWSB and generate the large top mass.

We have computed the corrections to the $Zb\bar b$ couplings at tree level by performing a perturbative diagonalization of the mass matrices expanding in powers of $\epsilon^2=\sin^2\left(\frac{v}{f_h}\right)$. We have obtained analytic expressions for $\delta g_{b_L}$ and $\delta g_{b_R}$ that allowed us to understand the origin, the sign and the size of the different contributions to these corrections. We have also studied the effective theory at energies below the scale of the heavy resonances. By integrating-out the {\it composite}-states we have obtained an effective Lagrangian for the {\it elementary}-states and the Higgs. In particular we have computed the $Zb\bar b$ and $Zt\bar t$ couplings using this procedure, which has the virtue that the symmetry properties of the different sectors become evident. 

Under certain assumptions, the class of models considered in our work has a finite Higgs potential at one-loop, that has been computed for fermions in different representations of SO(5). In the present article we have included in the calculation of the Higgs potential the contributions of all the fermions and vector bosons, including those arising from the bottom sector that transform with the representations ${\bf4}$ and ${\bf16}$, that can be important if the mixing between that quark and the heavy states are large. By performing a random scan we have found the regions of the parameter space where the EW symmetry is broken and the Higgs and the quarks of the third generation have the observed masses.  
We have also computed the full corrections to $Zb\bar b$ couplings at tree level numerically for those regions of the parameter space. We have checked that our analytic results are in very good agreement with the full numerical ones, with higher order corrections smaller than 1\% level. 

We have found a considerable region of the parameter space of our model where the $Zb\bar b$ couplings are corrected in such a way that they can solve the deviation attributed to $A^b_{FB}$. The model has been designed to produce those corrections, but we have found that generically, for {\it composite} bottom Yukawa couplings $y_b\sim g_\rho f_h$, the bottom mass and the correct shifts $\delta g_{b_L}$ and $\delta g_{b_R}$ can not be produced simultaneously. This is expected since there is a strong correlation between the shifts $\delta g_{b_L}$ and $\delta g_{b_R}$ arising from the experimental results. Instead we have found that $\delta g_{b_R}$ requires a rather large mixing of $b_R$ and its {\it composite} partners, as was expected, but also $\delta g_{b_L}$ requires a moderate mixing of $b_L$ and its partners. Since the shift $\delta g_{b_L}$ preferred by the experiments is small, one would have expected that a small mixing could be enough. However when going to the mass eigenstate basis, there is a correction to $g_{b_L}$ arising from the top sector, even in the case of $P_{LR}$-symmetry, that although being suppressed, has the wrong sign. Therefore, to compensate this correction, the $b_L$ mixing has to be somewhat larger than expected from a naive analysis. Since the mixing driving the corrections to $Zb\bar b$ also control the bottom mass, the {\it composite} bottom Yukawa has to be partially suppressed to lead to a light bottom. We have checked numerically that this is indeed the case, obtaining $y_b/f_h\sim{\cal O}(10^{-1})$, introducing a small hierarchy between $y_b$ and the other {\it composite} couplings $g_\rho$ that we take of order $g_{SM}\ll g_\rho\ll 4\pi$.

We have also estimated the corrections to the $Zb\bar b$ couplings at one-loop in the gaugeless limit. We found that these corrections are at least one order of magnitude smaller than the tree level ones. Therefore we can expect the tree level corrections to give a reliable estimation of the shifts of the $Zb\bar b$ couplings.

We found an interesting collider phenomenology involving the beautiful mirror fermions. Finding exotic fermions with charge $-4/3$ or $-7/3$ would provide good evidence for the solution we have presented. There is also a rich phenomenology involving neutral interactions, with the possibility of cascade decays mediated by $Z$- and $b$-resonances, and possible final states with several bottom quarks and $Z$ with large $p_T$.

\subsection*{Acknowledgements}

We thank Elisabetta Furlan and Jos\'e Santiago for clarifications on their calculation of the corrections to $Zb\bar b$ at one loop and Elisabetta Furlan for sharing with us her codes. We also thank Diptimoy Ghosh for useful discussions and for having pointed us to Ref. \cite{Ciuchini:2014dea}. 
L.D. wishes to thank the Department of Particle Physics and Astrophysics at Weizmann Institute of Sciences for hospitality during a stage of this work. L.D. and I.D. are partly supported by FONCYT-Argentina under the contract PICT-2013-2266 and U.N.Cuyo under the contract 2.013-C004.

\appendix
\section{Representations of SO(5)}\label{app-reps}
For a comprehensive description of the \textbf{4} and \textbf{5} representations of $SO(5)$, Refs.\cite{Agashe:2004rs} and \cite{Carena:2014ria} should be consulted. Here we will present the \textbf{16} representation of SO(5) in more detail. Given the size of the matrices needed to represent the generators of the algebra for this representation, and the fact that most of the elements of these matrices are zeroes, we will describe them by their diagonals. For that purpose, we define $\Diag(g,k)$ as the $k$-th diagonal of the matrix representing the element $g$ of the $so(5)$ algebra. $k = 0$ will be the main diagonal and we will use positive numbers to denote diagonals above the main one and negative numbers for those below it. As an example, if we consider the matrix, 
\begin{equation}
M_{ex} = \left[\begin{array}{ccc}
A1 & A2 & A3 \\
B1 & B2 & B3 \\
C1 & C2 & C3
\end{array}\right] \ ,
\end{equation}
then,
\begin{align}
&\Diag(M_{ex},0) = (A1,B2,C3) \ , \nonumber \\
&\Diag(M_{ex},1) = (A2,B3) \ , \qquad \Diag(M_{ex},2) = (A3) \ ,\nonumber \\
&\Diag(M_{ex},-1) = (B1,C2) \ , \quad \Diag(M_{ex},-2) = (C1) \ .
\end{align}

We now proceed to define the generators of the $so(5)$ algebra for the \textbf{16} as follows:
\begin{align}
&\Diag(T^{3}_{L},0)=\left(\frac{1}{2},-\frac{1}{2},1,0,0,-1,\frac{1}{2},\frac{1}{2},-\frac{1}{2},-\frac{1}{2},1,0,0,-1,\frac{1}{2},-\frac{1}{2}\right)& \nonumber\\
\nonumber\\
&\Diag(T^{3}_{R},0)=\left(1,1,\frac{1}{2},\frac{1}{2},\frac{1}{2},\frac{1}{2},0,0,0,0,-\frac{1}{2},-\frac{1}{2},-\frac{1}{2},-\frac{1}{2},-1,-1\right)& \nonumber\\
&\begin{cases}\Diag(T^{+}_{L},1)=& \left(\frac{1}{\sqrt{2}},0,1,0,0,0,0,0,0,0,0,0,1,0,\frac{1}{\sqrt{2}}\right)\\
\Diag(T^{+}_{L},2)=&
\left(0,0,0,1,0,0,-\frac{1}{\sqrt{2}},\frac{1}{\sqrt{2}},0,0,1,0,0,0\right)\end{cases}&\nonumber\\
&T^{-}_{L} = T^{+\dag}_{L}\nonumber&\\
&\begin{cases}\Diag(T^{+}_{R},6)=&\Big(-1,0,0,0,0,0,0,0,0,0\Big)\nonumber\\
\Diag(T^{+}_{R},7)=&\left(0,1,0,0,-\frac{1}{\sqrt{2}},0,0,0,-1\right)\nonumber\\
\Diag(T^{+}_{R},8)=&\left(0,0,\frac{1}{\sqrt{2}},0,0,\frac{1}{\sqrt{2}},1,0\right)\nonumber\\
\Diag(T^{+}_{R},9)=&\left(0,0,0,\frac{1}{\sqrt{2}},0,0,0\right)\nonumber
\end{cases}& \nonumber \\
&T^{-}_{R} = T^{+\dag}_{R}&\nonumber\\
&\begin{cases} \Diag(T^{+-},-4)=&\left(0,0,0,\frac{\sqrt{5}}{4},0,\frac{1}{2}\sqrt{\frac{5}{2}},-\frac{1}{2\sqrt{2}},0,\frac{1}{4},0,0,0\right)\nonumber\\
\Diag(T^{+-},-3)=&\left(0,-\frac{1}{2}\sqrt{\frac{5}{2}},0,\frac{1}{4},-\frac{3}{4},-\frac{1}{2\sqrt{2}},0,\frac{1}{2}\sqrt{\frac{5}{2}},\frac{\sqrt{5}}{4},\frac{\sqrt{5}}{4},0,-\frac{1}{2}\sqrt{\frac{5}{2}},0\right)\nonumber\\
\Diag(T^{+-},-2)=&\left(\frac{1}{2},\frac{1}{2\sqrt{2}},0,0,\frac{\sqrt{5}}{4},0,0,0,0,-\frac{3}{4},0,0,\frac{1}{2\sqrt{2}},\frac{1}{2}\right)\nonumber
\end{cases}&\\
&T^{-+} = T^{+-\dag}&\nonumber\\
&\begin{cases}\Diag(T^{++},3)=&\left(\frac{1}{2\sqrt{2}},0,0,0,0,0,0,0,0,0,0,0,-\frac{1}{2\sqrt{2}}\right)\nonumber\\
\Diag(T^{++},4)=&\left(\frac{1}{2}\sqrt{\frac{5}{2}},\frac{1}{2},-\frac{1}{2\sqrt{2}},0,-\frac{\sqrt{5}}{4},0,0,\frac{3}{4},0,\frac{1}{2}\sqrt{\frac{5}{2}},-\frac{1}{2},-\frac{1}{2}\sqrt{\frac{5}{2}}\right)\nonumber\\
\Diag(T^{++},5)=&\left(0,0,-\frac{1}{2}\sqrt{\frac{5}{2}},\frac{1}{4},-\frac{3}{4},0,\frac{\sqrt{5}}{4},\frac{\sqrt{5}}{4},\frac{1}{2\sqrt{2}},0,0\right)\nonumber\\
\Diag(T^{++},6)=&\left(0,0,0,-\frac{\sqrt{5}}{4},0,0,-\frac{1}{4},0,0,0\right)\nonumber\end{cases}&\\
&T^{--}=T^{++\dag}&
\end{align}
All elements not explicitly indicated are zeroes. The set $\{T^{3}_{L},T^{+}_{L},T^{-}_{L},T^{3}_{R},T^{+}_{R},T^{-}_{R}\}$ generates the $so(4)$ algebra that is left unbroken after the SO(5)$\rightarrow$SO(4) breaking. As is well known, the reason for choosing the usual $SU(2)$ naming convention for the generators of this sub-algebra is that SO(4)$\simeq$SU(2)$_L\times$SU(2)$_R$. On the other hand, $\{T^{+-},T^{-+},T^{++},T^{--}\}$ are the broken generators corresponding to the $SO(5)/SO(4)$ coset, that transform as a $({\bf2},{\bf2})$ of SO(4). The first (second) $\pm$ super-index of these generators stands for the $\pm1/2$ eigenvalue under $T^3_L$ ($T^3_R$).

The basis we have chosen to describe these matrices is a basis of eigenvectors of $T^{3}_{L}$ and $T^{3}_{R}$ (as can be readily seen by the fact that their matrices are diagonal) which are also eigenvectors of the Casimir operators, $(T^{1}_{L})^{2} + (T^{2}_{L})^{2} + (T^{3}_{L})^{2}$ and $(T^{1}_{R})^{2} + (T^{2}_{R})^{2} + (T^{3}_{R})^{2}$ \, \footnote{As usual, $T^{1}_{L/R} = T^{+}_{L/R} + T^{-}_{L/R}$ and $T^{1}_{L/R} =-i(T^{+}_{L/R} - T^{-}_{L/R})$}. Below we show our notation for the components of the fermions corresponding to the different representations of SO(5), with charges under SU(2)$_L\times$SU(2)$_R$ given by the previous generators:
\begin{align}
&\psi_{4} = \left(B_{1}, V_{1}, B, V_{2} \right) \ ,
\\
&\psi_{5} = \left(-\frac{i}{\sqrt{2}} (X - B), \frac{1}{\sqrt{2}}
(X + B), \frac{i}{\sqrt{2}} (T_{1} - T),\frac{1}{\sqrt{2}} (T_{1}
+ T), \tilde{T} \right) \ ,
\\
&\psi_{16} = \left(T, B, T_{1}, B_{1}, B_{2}, V_{1}, B_{3}, B_{4},
V_{2}, V_{3}, B_{5}, V_{4}, V_{5}, S_{1}, V_{6}, S_{2} \right) \ .
\end{align}
This notation will become useful to describe a basis for the mass and Yukawa matrices of appendix \ref{App-yukawa}.

\section{Yukawa interactions and mass matrices}\label{App-yukawa}
In this appendix we show the neutral Yukawa terms and the mass matrices for the up- and down-type fermions. The neutral Yukawa terms involving the ${\bf4}$ and ${\bf16}$ representations are:
\begin{eqnarray}\label{YukFullTerms}
\mathcal{L}^{b}_{\text{yuk}} &=& \frac{1}{8} y_{b({\bf 1},{\bf 2})}
\Big[-\Big(B^{B}_{R} \, c_{\frac{h}{2}} + i B^{B}_{1,R} \,
\, s_{\frac{h}{2}}\Big) \Big(-(3 \, c_{\frac{h}{2}} + 5
\, c_{\frac{3h}{2}}) \overline{B^{Q^b}_{2,L}} + \nonumber
\nonumber \\ &+& 2 \, s_{\frac{h}{2}} (- (1 + 5
\, c_{h}) \overline{B^{Q^b}_{4,L}} + \sqrt{5}
(\sqrt{2} \, ct_{\frac{h}{2}} \overline{B^{Q^b}_{L}} -
\overline{B^{Q^b}_{1,L}} + \, ct_{\frac{h}{2}}
\overline{B^{Q^b}_{3,L}} + \nonumber \\ &+& \sqrt{2} \,
\overline{B^{Q^b}_{5,L}}) \, s_{h})\Big) + \Big((3
\, c_{\frac{h}{2}} + 5 \, c_{\frac{3h}{2}})
\overline{V^{Q^b}_{4,L}} + 2 \, s_{\frac{h}{2}} (-(1 + 5
\, c_{h}) \overline{V^{Q^b}_{3,L}} + \nonumber \\ &+&
\sqrt{5} (-\sqrt{2} \, \overline{V^{Q^b}_{1,L}} +
\, ct_{\frac{h}{2}} \overline{V^{Q^b}_{2,L}} +
\overline{V^{Q^b}_{5,L}} + \sqrt{2} \, ct_{\frac{h}{2}}
\overline{V^{Q^b}_{6,L}}) \, s_{h})\Big)\Big(i
\, s_{\frac{h}{2}} V^{B}_{1,R} + \nonumber \\ &+&
\, c_{\frac{h}{2}} V^{B}_{2,R}\Big)\Big] + \frac{1}{8} y_{b({\bf 2},{\bf 1})}
\Big[\Big(B^{B}_{1,R} \, c_{\frac{h}{2}} + i B^{B}_{R}
\, s_{\frac{h}{2}}\Big)\Big((3 \, c_{\frac{h}{2}} + 5
\, c_{\frac{3h}{2}}) \overline{B^{Q^b}_{4,L}} + \nonumber \\
&+& 2 \, s_{\frac{h}{2}}(2 \sqrt{5} \, c^{2}_{\frac{h}{2}}
\overline{B^{Q^b}_{1,L}} - 2 \sqrt{10} \, c^{2}_{\frac{h}{2}}
\overline{B^{Q^b}_{5,L}} - \overline{B^{Q^b}_{2,L}} - 5
\, c_{h} \overline{B^{Q^b}_{2,L}} + \nonumber \\ &+&
\sqrt{10} \, \overline{B^{Q^b}_{L}} \, s_{h} +
\sqrt{5} \, \overline{B^{Q^b}_{3,L}} \, s_{h})\Big) +
\Big((3 \, c_{\frac{h}{2}} + 5 \, c_{\frac{3h}{2}})
\overline{V^{Q^b}_{3,L}} + \nonumber \\ &+& 2
\, s_{\frac{h}{2}}(2 \sqrt{10} \, c^{2}_{\frac{h}{2}}
\overline{V^{Q^b}_{1,L}} - 2 \sqrt{5} \, c^{2}_{\frac{h}{2}}
\overline{V^{Q^b}_{5,L}} + \overline{V^{Q^b}_{4,L}} + 5
\, c_{h} \overline{V^{Q^b}_{4,L}} + \nonumber \\
&+&  \sqrt{5} \, \overline{V^{Q^b}_{2,L}} \, s_{h} +
\sqrt{10} \, \overline{V^{Q^b}_{6,L}}
\, s_{h})\Big)\Big(\, c_{\frac{h}{2}} V^{B}_{1,R} + i
\, s_{\frac{h}{2}} V^{B}_{2,R}\Big)\Big] + {\rm h.c.}
\end{eqnarray}
In the last expression $ct_{n h} = \cot(\frac{n h}{f_{h}})$, and as usual $c_{n h} = \cos(\frac{n h}{f_h})$ and $s_{n h} = \sin(\frac{n h}{f_h})$. The neutral Yukawa terms involving the ${\bf5}$ representations are:
\begin{eqnarray}\label{topyuk}
\mathcal{L}^{t}_{\text{yuk}} &=& y_{t(\mathbf{1},\mathbf{1})}
\left[c_{h} \overline{\tilde{T}^{Q^t}_{L}} -
 \frac{s_{h}}{\sqrt{2}}(\overline{T^{Q^t}_{L}} \, + \,
 \overline{T^{Q^t}_{1,L}}) \right]
\left[ c_{h} \tilde{T}^{T}_{R} - \frac{s_{h}}{\sqrt{2}}(T^{T}_{R}
  + T^{T}_{1,R}) \right] \nonumber + \\
&+& y_{t(\mathbf{2},\mathbf{2})} \Big\{
\overline{B^{Q^t}_{L}} B^{T}_{R} +
\overline{X^{Q^t}_{L}} X^{T}_{R} +
\frac{1}{2} \left(\overline{T^{Q^t}_{L}} -
\overline{T^{Q^t}_{1,L}}\right) \left(T^{T}_{R} -
T^{T}_{1,R}\right) + \nonumber \\
&+& \left[\frac{c_{h}}{\sqrt{2}}\left(\overline{T^{Q^t}_{L}} +
\overline{T^{Q^t}_{1,L}}\right) + s_{h}
\overline{\tilde{T}^{Q^t}_{L}} \right]
\left[\frac{c_{h}}{\sqrt{2}}\left(\overline{T^{T}_{R}} +
\overline{T^{T}_{1,R}}\right) + s_{h} \overline{\tilde{T}^{T}_{R}}
\right] \Big\} \nonumber + \\ &+& {\rm h.c.}
\end{eqnarray}
We can calculate a $\overline{L}R$ matrix for the quadratic terms
of the $b$ sector at zero momentum from the Lagrangian, using the
basis
$\left\{{\overline{b_{L}},\overline{B^{B}_{1,L}},\overline{B^{B}_{L}},\overline{B^{Q^b}_{L}},\overline{B^{Q^b}_{1,L}},\overline{B^{Q^b}_{2,L}},\overline{B^{Q^b}_{3,L}},\overline{B^{Q^b}_{4,L}},\overline{B^{Q^b}_{5,L}},\overline{B^{T}_{L}},\overline{B^{Q^t}_{L}}}\right\}$
and
$\left\{{b_{R}},B^{B}_{1,R},B^{B}_{R},B^{Q^b}_{R},B^{Q^b}_{1,R},B^{Q^b}_{2,R},B^{Q^b}_{3,R},B^{Q^b}_{4,R},B^{Q^b}_{5,R},B^{T}_{R},B^{Q^t}_{R}\right\}$,

\begin{equation}
M^{b}_{\overline{L}R} = \left[\begin{array}{cccccccc} 0 & 0 & 0 & \Delta_{Q^b} & 0 & \cdots & 0 & \Delta_{Q^t} \\
0 & -m_{B} & 0 & 0 & 0 & \cdots & 0 & 0 \\
\Delta b & 0 & -m_{B} & 0 & 0 & \cdots & 0 & 0 \\
0 & \multicolumn{2}{c}{\multirow{2}{*}{\Bigg($Y^{b}_{6\times2}$\Bigg)}} & \multicolumn{2}{c}{\multirow{2}{*}{$\, \, \, \, \, \, \ddots$}} &  & \multirow{2}{*}{$\, \vdots$} & \multirow{2}{*}{$\, \vdots$}\\
\vdots & & & & & & & \\
0 & 0 & 0 & 0 & \cdots & & -m_{T} & 0 \\
0 & 0 & 0 & 0 & \cdots & & y_{t(\mathbf{2},\mathbf{2})} & -m_{Q^{t}}
\end{array}\right] \ .
\end{equation}
where the main diagonal is $(0,-m_{B},-m_{B},-m_{Q^{b}},-m_{Q^{b}},-m_{Q^{b}},-m_{Q^{b}},-m_{Q^{b}},-m_{Q^{b}},-m_{T},-m_{Q^{t}})$ and the Yukawa sub-matrix for the $b$ sector, $Y^{b}_{6 \times 2}$, is
\begin{equation}
Y^{b}_{6 \times 2} =
\scalebox{.75}{$
 \frac{1}{4} \left[\begin{array}{cc} \sqrt{\frac{5}{2}} (- i y_{b({\bf 1},{\bf 2})} + y_{b({\bf 2},{\bf 1})}) s^{2}_{h} & -\sqrt{\frac{5}{2}} (y_{b({\bf 1},{\bf 2})} - i y_{b({\bf 2},{\bf 1})} + (y_{b({\bf 1},{\bf 2})} + i y_{b({\bf 2},{\bf 1})}) c_{h}) s_{h} \\
\frac{\sqrt{5}}{2}(i y_{b({\bf 1},{\bf 2})} + y_{b({\bf 2},{\bf 1})} + (-i y_{b({\bf 1},{\bf 2})} +
y_{b({\bf 2},{\bf 1})}) c_{h})s_{h} & \frac{\sqrt{5}}{2} (y_{b({\bf 1},{\bf 2})}+i
y_{b({\bf 2},{\bf 1})})s^{2}_{h}
\\ \frac{i}{2} (- y_{b({\bf 1},{\bf 2})} + i y_{b({\bf 2},{\bf 1})} + 5 (y_{b({\bf 1},{\bf 2})} + i y_{b({\bf 2},{\bf 1})}) c_{h}) s_{h} & \frac{1}{4}(8(y_{b({\bf 1},{\bf 2})}- i y_{b({\bf 2},{\bf 1})})c_{h} + (y_{b({\bf 1},{\bf 2})} + i y_{b({\bf 2},{\bf 1})}) (3 + 5 c_{2h})) \\
\frac{\sqrt{5}}{2}(-i y_{b({\bf 1},{\bf 2})} + y_{b({\bf 2},{\bf 1})}) s^{2}_{h} & -\frac{\sqrt{5}}{2}(y_{b({\bf 1},{\bf 2})}-i y_{b({\bf 2},{\bf 1})} + (y_{b({\bf 1},{\bf 2})}+ i y_{b({\bf 2},{\bf 1})})c_{h})s_{h} \\
\frac{1}{4}(8(i y_{b({\bf 1},{\bf 2})} + y_{b({\bf 2},{\bf 1})})c_{h} + (-i y_{b({\bf 1},{\bf 2})} + y_{b({\bf 2},{\bf 1})}) (3 + 5 c_{2h})) & \frac{1}{2} (y_{b({\bf 1},{\bf 2})} - i y_{b({\bf 2},{\bf 1})} + 5 (y_{b({\bf 1},{\bf 2})} + i y_{b({\bf 2},{\bf 1})}) c_{h}) s_{h} \\
-\sqrt{\frac{5}{2}} (i y_{b({\bf 1},{\bf 2})} + y_{b({\bf 2},{\bf 1})} + (-i y_{b({\bf 1},{\bf 2})} +
y_{b({\bf 2},{\bf 1})}) c_{h}) s_{h} & -\sqrt{\frac{5}{2}} (y_{b({\bf 1},{\bf 2})} + i
y_{b({\bf 2},{\bf 1})}) s^{2}_{h}
\end{array}\right] \ .
$}
\end{equation}

In a similar fashion, using the basis
$\left\{{\overline{t_{L}},\overline{T^{T}_{L}},\overline{T^{T}_{1,L}},\overline{\tilde{T}^{T}_{L}},\overline{T^{Q^b}_{L}},\overline{T^{Q^b}_{1,L}},\overline{T^{Q^t}_{L}},\overline{T^{Q^t}_{1,L}},\overline{\tilde{T}^{Q^t}_{L}}}\right\}$
and \\
$\left\{{t_{R}},T^{T}_{R},T^{T}_{1,R},\tilde{T}^{T}_{R},T^{Q^b}_{R},T^{Q^b}_{1,R},T^{Q^t}_{R},T^{Q^t}_{1,R},\tilde{T}^{Q^t}_{R}\right\}$,
we obtain for the top sector:
\begin{equation}
M^{t}_{\overline{L}R} = \left[\begin{array}{ccccccccc} 0 & 0 & 0 & 0 & \Delta_{Q^{b}} & 0 & - \Delta_{Q^{t}} & 0 & 0 \\
0 & -m_{T} & 0 & 0 & 0 & 0 & 0 & 0 & 0 \\
0 & 0 & -m_{T} & 0 & 0 & 0 & 0 & 0 & 0 \\
\Delta_{T} & 0 & 0 & -m_{T} & 0 & 0 & 0 & 0 & 0\\
0 & 0 & 0 & 0 & -m_{Q^{b}} & 0 & 0 & 0 & 0 \\
0 & 0 & 0 & 0 & 0 & -m_{Q^{b}} & 0 & 0 & 0 \\
0 & \multicolumn{3}{c}{\multirow{3}{*}{\Bigg($Y^{t}_{3\times3}$\Bigg)}} & 0 & 0 & -m_{Q^{t}} & 0 & 0 \\
0 &  &  &  & 0 & 0 & 0 & -m_{Q^{t}} & 0 \\
0 &  &  &  & 0 & 0 & 0 & 0 & -m_{Q^{t}} \\
\end{array}\right] \ .
\end{equation}
where $Y^{t}_{3 \times 3}$ is given by:
\begin{equation}
Y^{t}_{3 \times 3} = \left[\begin{array}{ccc} \frac{1}{2} \left(y_{t(\mathbf{2},\mathbf{2})} (1 +  c^{2}_{h}) + y_{t(\mathbf{1},\mathbf{1})} \, s^{2}_{h}\right) & \frac{1}{2} \left(y_{t(\mathbf{1},\mathbf{1})} - y_{t(\mathbf{2},\mathbf{2})}\right) \, s^{2}_{h} & \left(y_{t(\mathbf{2},\mathbf{2})} - y_{t(\mathbf{1},\mathbf{1})}\right) \, \frac{c_{h} \, s_{h}}{\sqrt{2}} \\
\frac{1}{2} \left(y_{t(\mathbf{1},\mathbf{1})} -
y_{t(\mathbf{2},\mathbf{2})}\right) \, s^{2}_{h} & \frac{1}{2}
\left(y_{t(\mathbf{2},\mathbf{2})} (1 +  c^{2}_{h}) +
y_{t(\mathbf{1},\mathbf{1})} \, s^{2}_{h}\right) &
\left(y_{t(\mathbf{2},\mathbf{2})} - y_{t(\mathbf{1},\mathbf{1})}\right) \,\frac{c_{h} \, s_{h}}{\sqrt{2}} \\
\left(y_{t(\mathbf{2},\mathbf{2})} -
y_{t(\mathbf{1},\mathbf{1})}\right) \,\frac{c_{h} \,
s_{h}}{\sqrt{2}} & \left(y_{t(\mathbf{2},\mathbf{2})} -
y_{t(\mathbf{1},\mathbf{1})}\right) \, \frac{c_{h} \,
s_{h}}{\sqrt{2}} & y_{t(\mathbf{1},\mathbf{1})} \, c^{2}_{h} +
y_{t(\mathbf{2},\mathbf{2})} \, s^{2}_{h}
\end{array}\right] \ .
\end{equation}

Now, for the exotic quarks in the model, we begin by considering the $v$-type quarks with the basis
$\left\{\overline{V^{B}_{1,L}},\overline{V^{B}_{2,L}},\overline{V^{Q^b}_{1,L}},\overline{V^{Q^b}_{2,L}},\overline{V^{Q^b}_{3,L}},\overline{V^{Q^b}_{5,L}},\overline{V^{Q^b}_{4,L}},\overline{V^{Q^b}_{6,L}}\right\}$
and
$\left\{V^{B}_{1,R},V^{B}_{2,R},V^{Q^b}_{1,R},V^{Q^b}_{2,R},V^{Q^b}_{3,R},V^{Q^b}_{5,R},V^{Q^b}_{4,R},V^{Q^b}_{6,R}\right\}$,
and obtain:
\begin{equation}
M^{v}_{\overline{L}R} = \left[\begin{array}{cccc} -m_{B} & 0 & \cdots & 0 \\
0 & -m_{B} & \cdots & 0 \\
\multicolumn{2}{c}{\multirow{2}{*}{\Bigg($Y^{v}_{6\times2}$\Bigg)}} & \ddots & \vdots\\
 & & & -m_{Q^{b}}
\end{array}\right] \ .
\end{equation}
where the main diagonal is
$(-m_{B},-m_{B},-m_{Q^{b}},-m_{Q^{b}},-m_{Q^{b}},-m_{Q^{b}},-m_{Q^{b}},-m_{Q^{b}})$
and all off-diagonal elements are zero except for the $6 \times 2$ sub-matrix $Y^{v}_{6\times2}$,
\begin{equation}
Y^{v}_{6 \times 2} =
\scalebox{.74}{$
\frac{1}{4} \left[\begin{array}{cc}
\sqrt{\frac{5}{2}}(-i y_{b({\bf 1},{\bf 2})} +
y_{b({\bf 2},{\bf 1})} + (i y_{b({\bf 1},{\bf 2})} +
y_{b({\bf 2},{\bf 1})}) c_{h})s_{h} &  -\sqrt{\frac{5}{2}} (y_{b({\bf 1},{\bf 2})} - i y_{b({\bf 2},{\bf 1})}) s^{2}_{h} \\
\frac{\sqrt{5}}{2} (i y_{b({\bf 1},{\bf 2})} +
y_{b({\bf 2},{\bf 1})})s^{2}_{h} &
\frac{\sqrt{5}}{2}(y_{b({\bf 1},{\bf 2})} + i
y_{b({\bf 2},{\bf 1})} + (y_{b({\bf 1},{\bf 2})} - i
y_{b({\bf 2},{\bf 1})}) c_{h})s_{h}
\\ \frac{1}{4}(8(-i y_{b({\bf 1},{\bf 2})} + y_{b({\bf 2},{\bf 1})})c_{h} + (i y_{b({\bf 1},{\bf 2})} + y_{b({\bf 2},{\bf 1})}) (3 + 5 \, c_{2h})) & -\frac{1}{2} (y_{b({\bf 1},{\bf 2})} + i y_{b({\bf 2},{\bf 1})} + 5 (y_{b({\bf 1},{\bf 2})} - i y_{b({\bf 2},{\bf 1})}) c_{h}) s_{h} \\
-\frac{\sqrt{5}}{2}(-i y_{b({\bf 1},{\bf 2})} + y_{b({\bf 2},{\bf 1})} + (i y_{b({\bf 1},{\bf 2})}+ y_{b({\bf 2},{\bf 1})})c_{h})s_{h} & \frac{\sqrt{5}}{2}(y_{b({\bf 1},{\bf 2})} - i y_{b({\bf 2},{\bf 1})}) s^{2}_{h} \\
\frac{1}{2} (-i y_{b({\bf 1},{\bf 2})} + y_{b({\bf 2},{\bf 1})} + 5 (i y_{b({\bf 1},{\bf 2})} + y_{b({\bf 2},{\bf 1})}) c_{h}) s_{h} & \frac{1}{4}(8(y_{b({\bf 1},{\bf 2})} + i y_{b({\bf 2},{\bf 1})})c_{h} + (y_{b({\bf 1},{\bf 2})} - i y_{b({\bf 2},{\bf 1})}) (3 + 5 c_{2h})) \\
\sqrt{\frac{5}{2}} (i y_{b({\bf 1},{\bf 2})} +
y_{b({\bf 2},{\bf 1})}) s^{2}_{h} & \sqrt{\frac{5}{2}}
(y_{b({\bf 1},{\bf 2})} + i y_{b({\bf 2},{\bf 1})} +
(y_{b({\bf 1},{\bf 2})} - i y_{b({\bf 2},{\bf 1})})
c_{h}) s_{h}
\end{array}\right] \ .
$}
\end{equation}

For the $s$-type (charge -7/3) quarks, using the basis
$\left\{\overline{S^{Q^b}_{1,L}},\overline{S^{Q^b}_{2,L}}\right\}$
and $\left\{S^{Q^b}_{1,R},S^{Q^b}_{2,R}\right\}$, we get:
\begin{equation}
M^{s}_{\overline{L}R} = \left[\begin{array}{cc} -m_{Q^{b}} & 0 \\
0 & -m_{Q^{b}}
\end{array}\right] \ .
\end{equation}

And finally, for the $x$-type (charge 5/3) quarks, we use the basis
$\left\{\overline{X^{Q^t}_{L}}, \overline{X^{T}_{L}}\right\}$
and $\left\{X^{Q^t}_{R}, X^{T}_{R}\right\}$ and obtain:
\begin{equation}
M^{x}_{\overline{L}R} = \left[\begin{array}{cc} -m_{Q^{t}} & y_{t(\mathbf{2},\mathbf{2})} \\
0 & -m_{T}
\end{array}\right] \ .
\end{equation}

\section{Correlators}\label{App-correlators}
In this appendix we show the correlators obtained after integration of the {\it composite} states of our model at tree level. The correlators obtained from integration of the heavy vector bosons are:
\begin{align}
\Pi_{({\bf3},{\bf1})}^A = \Pi_{({\bf1},{\bf3})}^A = \frac{p^2 m^2_\rho}{g^2_\rho (p^2 - m^2_\rho)}~,
\hspace{1cm}
&\Pi_{({\bf2},{\bf2})}^A = \frac{m^2_\rho (p^2 + m^2_\rho - m^2_{\hat{a}})}{g^2_\rho (p^2 - m^2_{\hat{a}})}~,
\hspace{1cm}
\Pi^X = \frac{p^2 m^2_X}{g^2_X (p^2 - m^2_X)}~. \nonumber \\
&\Pi_{({\bf3},{\bf1})+({\bf1},{\bf3})}^A \equiv \frac{\Pi_{({\bf3},{\bf1})}^A + \Pi_{({\bf1},{\bf3})}^A}{2}.
\end{align}

The correlators arising from integration of $Q^t$ and $T$ are:
\begin{align}
\Pi^{q^t}_{(2,2)} &= \Delta_{Q^t}^{2} \frac{m_{T}^{2} - p^{2}
+ y_{t(\mathbf{2},\mathbf{2})}^2}{m_{Q^t}^{2}(m_{T}^{2} - p^{2}) + p^{2} (p^{2} -
  m_{T}^{2} - y_{t(\mathbf{2},\mathbf{2})}^2)} \, ,\nonumber \\
\Pi^{q^t}_{(1,1)} &= 
\Delta_{Q^t}^{2} \frac{m_{T}^{2} - p^{2} +
y_{t(\mathbf{1},\mathbf{1})}^2}{m_{Q^t}^{2}(m_{T}^{2} - p^{2}) + p^{2} (p^{2} -
m_{T}^{2} - y_{t(\mathbf{1},\mathbf{1})}^{2})} \nonumber \\
\Pi^t_{(2,2)} &= \Delta_{T}^{2} \frac{m_{Q^t}^{2} - p^{2} +
y_{t(\mathbf{2},\mathbf{2})}^2}{m_{Q^t}^{2}(m_{T}^{2} - p^{2}) + p^{2} (p^{2} -
  m_{T}^{2} - y_{t(\mathbf{2},\mathbf{2})}^{2})} \, ,\nonumber \\
\Pi^t_{(1,1)} &= \Delta_{T}^{2} \frac{m_{Q^t}^{2} - p^{2} +
y_{t(\mathbf{1},\mathbf{1})}^2}{m_{Q^t}^{2}(m_{T}^{2} - p^{2}) + p^{2} (p^{2} -
m_{T}^{2} - y_{t(\mathbf{1},\mathbf{1})}^{2})} \, ,\nonumber \\
M^t_{(2,2)} &= \Delta_{Q^t} \Delta_{T} \frac{m_{Q^t} \, m_{T} \,
y_{t(\mathbf{2},\mathbf{2})}}{m_{Q^t}^{2}(m_{T}^{2} - p^{2}) + p^{2} (p^{2} -
  m_{T}^{2} - y_{t(\mathbf{2},\mathbf{2})}^{2})} \, ,\nonumber \\
M^t_{(1,1)} &= \Delta_{Q^t}
\Delta_{T} \frac{m_{Q^t} \, m_{T} \, y_{t(\mathbf{1},\mathbf{1})}}{m_{Q^t}^{2}(m_{T}^{2} -
p^{2}) + p^{2} (p^{2} - m_{T}^{2} - y_{t(\mathbf{1},\mathbf{1})}^{2})} \nonumber \ .
\end{align}
The correlators arising from integration of $Q^b$ and $B$ are:
\begin{align}
\Pi^{q^b}_{(2,3)} &= \Pi^{q^b}_{(3,2)} = \frac{\Delta_{Q^b}^{2}} {m_{Q^b}^{2} - p^{2}} ,\nonumber\\
\Pi^{q^b}_{(2,1)} &= \Delta_{Q^b}^{2} \frac{m_{B}^{2} - p^{2}
+ y_{b({\bf 2},{\bf 1})}^2}{m_{B}^{2}(m_{Q^b}^{2} - p^{2}) + p^{2} (p^{2} -
  m_{Q^b}^{2} - y_{b({\bf 2},{\bf 1})}^{2})} \, ,\nonumber\\
\Pi^{q^b}_{(1,2)} &=
\Delta_{Q^b}^{2} \frac{m_{B}^{2} - p^{2} +
y_{b({\bf 1},{\bf 2})}^2}{m_{B}^{2}(m_{Q^b}^{2} - p^{2}) + p^{2} (p^{2} -
m_{Q^b}^{2} - y_{b({\bf 1},{\bf 2})}^{2})} \ ,\nonumber \\
\Pi^b_{(2,1)} &= \Delta_{B}^{2} \frac{m_{Q^b}^{2} - p^{2} +
y_{b({\bf 2},{\bf 1})}^2}{m_{B}^{2}(m_{Q^b}^{2} - p^{2}) + p^{2} (p^{2} -
  m_{Q^b}^{2} - y_{b({\bf 2},{\bf 1})}^{2})} \, ,\nonumber\\
\Pi^b_{(1,2)} &= \Delta_{B}^{2} \frac{m_{Q^b}^{2} - p^{2} +
y_{b({\bf 1},{\bf 2})}^2}{m_{B}^{2}(m_{Q^b}^{2} - p^{2}) + p^{2} (p^{2} -
m_{Q^b}^{2} - y_{b({\bf 1},{\bf 2})}^{2})} \ ,\nonumber \\
M^b_{(2,1)} &= \Delta_{Q^b} \Delta_{B} \frac{m_{Q^b} \, m_{b} \,
y_{b({\bf 2},{\bf 1})}}{m_{B}^{2}(m_{Q^b}^{2} - p^{2}) + p^{2} (p^{2} -
  m_{Q^b}^{2} - y_{b({\bf 2},{\bf 1})}^{2})} \, ,\nonumber\\
M^b_{(1,2)} &= \Delta_{Q^b}
\Delta_{B} \frac{m_{Q^b} \, m_{b} \, y_{b({\bf 1},{\bf 2})}}{m_{B}^{2}(m_{Q^b}^{2} -
p^{2}) + p^{2} (p^{2} - m_{Q^b}^{2} - y_{b({\bf 1},{\bf 2})}^{2})} \ .\nonumber
\end{align}


\end{document}